\documentstyle[12pt,epsf]{article}
\def\hybrid{\topmargin -20pt    \oddsidemargin 0pt
        \headheight 0pt \headsep 0pt
        \textwidth 6.5in        
        \textheight 9in         
        \marginparwidth .875in
        \parskip 5pt plus 1pt   \jot = 1.5ex}

\hybrid
\makeatletter
\@addtoreset{equation}{section}
\makeatother

\newcommand{\cC}{{\cal C}}

\newcommand{\cK}{{\cal K}}

\newcommand{\cM}{{\cal M}}
\newcommand{\cN}{{\cal N}}

\newcommand{\cP}{{\cal P}}

\newcommand{\cR}{{\cal R}}

\newcommand{\bea}{\begin{eqnarray}}
\newcommand{\eea}{\end{eqnarray}}
\newcommand{\be}{\begin{equation}}
\newcommand{\ee}{\end{equation}}
\newcommand{\bt}{\begin{tabular}}
\newcommand{\et}{\end{tabular}}
\newcommand{\ba}{\begin{array}}
\newcommand{\ea}{\end{array}}

\newcommand{\bmat}{\left(\begin{array}}
\newcommand{\emat}{\end{array}\right)}

\newcommand{\Tr}{\mathop{\rm Tr}}
\newcommand{\tr}{\mathop{\rm tr}}

\def\beq{\begin{equation}}
\def\eeq{\end{equation}}
\def\beqa{\begin{eqnarray}}
\def\eeqa{\end{eqnarray}}

\def\NPB#1#2#3{Nucl.\ Phys.\ B{#1} (19#2) #3}
\def\PLB#1#2#3{Phys.\ Lett.\ B{#1} (19#2) #3}

\def\PRD#1#2#3{Phys.\ Rev.\ D{#1} (19#2) #3}

\def\bfm#1{\mbox{\boldmath$#1$}}

\def\N#1{$\cN=#1$}
\def\D#1{$D=#1$}

\def\yzero{\smash{\hbox{$y\kern-4pt\raise1pt\hbox{${}^\circ$}$}}}

\def\s2{\frac{1}{\sqrt2}}

\def\bZ{{\mathbf Z}}

\def\half{\frac12}
\def\th{^{\rm th}}

\def\ie{i.e.\ }
\def\eg{e.g.\ }

\def\IB{\relax{\rm I\kern-.18em B}}
\def\ID{\relax{\rm I\kern-.18em D}}
\def\IE{\relax{\rm I\kern-.18em E}}
\def\IF{\relax{\rm I\kern-.18em F}}
\def\IH{\relax{\rm I\kern-.18em H}}
\def\II{\relax{\rm I\kern-.18em I}}
\def\IK{\relax{\rm I\kern-.18em K}}
\def\IL{\relax{\rm I\kern-.18em L}}
\def\IM{\relax{\rm I\kern-.18em M}}
\def\IN{\relax{\rm I\kern-.18em N}}
\def\IP{\relax{\rm I\kern-.18em P}}
\def\IR{\relax{\rm I\kern-.18em R}}
\def\IT{\relax{\rm I\kern-.42em T}}
\def\IZ{\relax{\hbox{\raisebox{.38ex}
    {\scriptsize\bfseries\slshape /}\kern-.40em\_\kern-.28em\rm Z}}}
\def\Iz{\relax{\hbox{\raisebox{.38ex}
    {\tiny\bfseries\slshape /}\kern-.25em\raisebox{.65ex}
    {\tiny\bfseries\slshape /}\kern-.43em\_\kern-.26em\rm Z}}}
\def\inbar{\vrule height1.5ex width.8pt depth-0.2pt}
\def\inbarhi{\vrule height1.55ex width.5pt depth-.85ex}
\def\inbarlo{\vrule height.8ex width.5pt depth0ex}
\def\IC{\relax{\rm C\kern-.48em \inbar\kern.48em}}
\def\IO{\relax{\rm O\kern-.56em \inbar\kern.56em}}
\def\IQ{\relax{\rm Q\kern-.56em \inbar\kern.56em}}
\def\IS{\relax{\rm S\kern-.37em \inbarhi\kern.08em\inbarlo\kern.29em}}

\def \one{\relax{\rm 1\kern-.26em I}}

\def\Dsl{\,\raise.15ex\hbox{/}\mkern-13.5mu D} 

\def\cp#1{\relax\ifmmode {\IP\kern-2pt{}_{#1}}\else $\IP\kern-2pt{}_{#1}$\fi}

\newcommand{\drawsquare}[2]{\hbox{%
\rule{#2pt}{#1pt}\hskip-#2pt
\rule{#1pt}{#2pt}\hskip-#1pt
\rule[#1pt]{#1pt}{#2pt}}\rule[#1pt]{#2pt}{#2pt}\hskip-#2pt
\rule{#2pt}{#1pt}}

\newcommand{\Yfun}{{\raisebox{-.5pt}{\drawsquare{6.5}{0.4}}}}
\newcommand{\Ysym}{{\raisebox{-.5pt}{\drawsquare{6.5}{0.4}}\hskip-0.4pt%
     \raisebox{-.5pt}{\drawsquare{6.5}{0.4}}}}
\newcommand{\Yasym}{{\raisebox{-3.5pt}{\drawsquare{6.5}{0.4}}\hskip-6.9pt%
     \raisebox{3pt}{\drawsquare{6.5}{0.4}}}}
\def \Yfunb{\overline\Yfun}
\def \Ysymb{\overline\Ysym}
\def \Yasymb{\overline\Yasym}

\def\cnodea{\put(0.8,2.5){\circle*{1.6}}}
\def\cnodeb{\put(7.4,2.5){\circle*{1.6}}}
\def\rnodea{\put(0.8,2.5){\circle{1.6}}}
\def\rnodeb{\put(7.4,2.5){\circle{1.6}}}
\def\pnodea{\put(0.8,2.5){\circle{1.6}}\put(0.8,2.5){\circle*{0.4}}}
\def\pnodeb{\put(7.4,2.5){\circle{1.6}}\put(7.4,2.5){\circle*{0.4}}}
\newlength{\firstlength} \newlength{\secondlength}
\def\qlink#1#2{\begin{picture}(9,4)
            \def\first{#1} \def\second{#2}
            \settowidth{\firstlength}{$k$} \settowidth{\secondlength}{$l$}
            \addtolength{\firstlength}{-0.5\firstlength}
            \addtolength{\secondlength}{-0.5\secondlength}
            \def\cx{c} \def\rx{r} \def\px{p}
            \ifx\first\cx \cnodea \else
               \ifx\first\rx \rnodea \else \pnodea \fi\fi
            \ifx\second\cx \put(1.6,2.5){\vector(1,0){5}}\cnodeb \else
               \put(1.6,2.5){\line(1,0){5}} \ifx\second\rx \rnodeb \else
               \pnodeb\fi\fi
            \put(1,-0.9){\hspace{-\firstlength}\scriptsize$k$}
            \put(7.6,-0.9){\hspace{-\secondlength}\scriptsize$l$}
            \end{picture}}
\def\qlinkx#1#2#3#4#5#6{\begin{picture}(9,4)
            \def\tail{#1} \def\head{#2}
            \def\first{#3} \def\second{#4}
            \settowidth{\firstlength}{$#5$} \settowidth{\secondlength}{$#6$}
            \addtolength{\firstlength}{-0.5\firstlength}
            \addtolength{\secondlength}{-0.5\secondlength}
            \def\cx{c} \def\rx{r} \def\px{p} \def\yes{1} \def\no{0}
            \ifx\first\cx \cnodea \else
               \ifx\first\rx \rnodea \else \pnodea \fi\fi
            \if\head\yes \put(1.6,2.5){\vector(1,0){5}} 
               \else\ifx\tail\no \put(1.6,2.5){\line(1,0){5}} \fi\fi
            \if\tail\yes \put(6.6,2.5){\vector(-1,0){5}} \fi
            \ifx\second\cx \cnodeb \else
               \ifx\second\rx \rnodeb \else \pnodeb\fi\fi
            \put(1,-0.9){\hspace{-\firstlength}\scriptsize$#5$}
            \put(7.6,-0.9){\hspace{-\secondlength}\scriptsize$#6$}
            \end{picture}}
\def\qtens#1#2{\def\first{#1}\def\cx{c}\def\rx{r}\def\px{p}%
               \settowidth{\firstlength}{$#2$}
               \addtolength{\firstlength}{-0.5\firstlength}
             \ifx\first\cx 
                   \begin{picture}(3,6)
                     \cnodea
                     \put(0.8,4.5){\circle{4}}
                     \put(1,-0.9){\hspace{-\firstlength}\scriptsize$#2$}
                   \end{picture}%
              \else%
                   \begin{picture}(3,6)
                     \ifx\first\rx\rnodea\else\pnodea\fi
                     \qbezier(-0.1,3)(-1.2,3.72843)(-1.2,4.5)
                     \qbezier(-1.2,4.5)(-1.2,5.32843)(-0.61421,5.91421)
                     \qbezier(-0.61421,5.91421)(-0.02843,6.5)(0.8,6.5)
                     \qbezier(0.8,6.5)(1.62843,6.5)(2.21421,5.91421)
                     \qbezier(2.21421,5.91421)(2.8,5.32843)(2.8,4.5)
                     \qbezier(2.8,4.5)(2.8,3.6)(1.6,3)
                     \put(1,-0.9){\hspace{-\firstlength}\scriptsize$#2$}
                   \end{picture}%
              \fi}
\def\qconjtens#1{\begin{picture}(3,6)
                \settowidth{\firstlength}{$#1$}
                \addtolength{\firstlength}{-0.5\firstlength}
                \cnodea
                \put(0.8,4.5){\circle{4}}
                \put(-1.2,3.8){\vector(1,-1){1}}
                \put(2.8,3.8){\vector(-1,-1){1}}
                \put(1,-0.9){\hspace{-\firstlength}\scriptsize$#1$}
                \end{picture}}
\def\qadj#1{\begin{picture}(3,6)
            \settowidth{\firstlength}{$#1$}
            \addtolength{\firstlength}{-0.5\firstlength}
            \cnodea
            \put(0.8,4.5){\circle{4}}
            \put(2.8,3.8){\vector(-1,-1){1}}
            \put(1,-0.9){\hspace{-\firstlength}\scriptsize$#1$}
            \end{picture}}

\def \nonsusy{non\discretionary{-}{}{-}super\-sym\-met\-ric\ }

\def \nonab{non\discretionary{-}{}{-}Abel\-ian\ }
\def \nonvan{non\discretionary{-}{}{-}van\-ish\-ing\ }
\def \noneq{non\discretionary{-}{}{-}equi\-va\-lent\ }

\def \noncom{non\discretionary{-}{}{-}com\-pact\ }

\begin{document}

\pagestyle{empty}
\renewcommand{\thefootnote}{\fnsymbol{footnote}}
\rightline{FTUAM-00/15, IFT-UAM/CSIC-00-22}
\rightline{\tt hep-th/0007087}
\vspace{0.5cm}
\begin{center}
\LARGE{\bf $D=4$, $\cN=1$ orientifolds with vector structure\\[20mm]}

\large{M.~Klein, R.~Rabad\'an\\[5mm]}

\small{Departamento de F\'{\i}sica Te\'orica C-XI
       and Instituto de F\'{\i}sica Te\'orica  C-XVI,\\[-0.3em]
       Universidad Aut\'onoma de Madrid,
       Cantoblanco, 28049 Madrid, Spain.\footnote{E-mail: 
       matthias.klein@uam.es, rabadan@delta.ft.uam.es}\\[20mm]}

\small{\bf Abstract} \\[7mm]
\end{center}

\begin{center}
\begin{minipage}[h]{14.0cm}

{\small
We construct compact type IIB orientifolds with discrete groups $\bZ_4$, 
$\bZ_6$, $\bZ_6'$, $\bZ_8$, $\bZ_{12}$ and $\bZ_{12}'$. These 
models are $\cN=1$ supersymmetric in $D=4$ and have vector structure. 
The possibility of having vector structure in $\bZ_N$ orientifolds with 
even $N$ arises due to an alternative $\Omega$-projection in the twisted 
sectors. Some of the models without vector structure are known to be 
inconsistent because of uncancelled tadpoles. We show that vector 
structure leads to a sign flip in the twisted Klein bottle contribution. 
As a consequence, all the tadpoles can be cancelled by introducing 
$D9$-branes and $D5$-branes. }

\end{minipage}
\end{center}
\newpage

\setcounter{page}{1}
\pagestyle{plain}
\renewcommand{\thefootnote}{\arabic{footnote}}
\setcounter{footnote}{0}


\section{Introduction}

Four-dimensional \N1 supersymmetric compact type IIB orientifold models have 
been extensively studied in the past \cite{bl,abpss,ks1,ks2,z,afiv,kst,kcpw,
bgk}. The usual construction is based on a product of the orbifold group 
$\Gamma$ and the world sheet parity $\Omega$ such that the whole orientifold 
group is of the form: $\Gamma + \Omega\Gamma$. The action of the  orientifold 
group on the Chan-Paton matrices is specified by choosing a projective real 
representation of the orbifold group $\Gamma$ \cite{bi,kr1}. It is known 
that there is a freedom in the choice of this representation, whenever 
$\Gamma$ contains elements of even order (\ie the smallest positive integer 
$N$, such that $g^N=e$, is even for some $g\in\Gamma$, where $e$ is the
neutral element of $\Gamma$). This freedom is related to the notion of 
vector structure in orientifold models, defined by the authors of 
\cite{blpssw}. It can also be understood from the theory of projective
representations of finite groups (see \eg \cite{kr1}). For all $\Gamma$
with only elements of odd order, one can always represent an element $g$
satisfying $g^N=e$ by a matrix $\gamma_g$ satisfying $\gamma_g^N=\one$.
But if $\Gamma$ contains elements of even order, there are two inequivalent
choices for the representation matrix $\gamma_g$ of an element $g$ of even
order $N$: $\gamma_g^N = \mu\,\one$, with $\mu=\pm1$. Orientifold models with
$\mu=+1$ ($\mu=-1$) have been called to have (no) vector structure in
\cite{blpssw}.   

The six-dimensional $\IZ_2$ orientifold first constructed by Bianchi and
Sagnotti \cite{bs} (their fourth example) and later analysed in detail by 
Gimon and Polchinski \cite{gp} has no vector structure.\footnote{Note that 
in \cite{gp} the authors choose a basis such that $\gamma_R^2=\one$, but 
$\gamma_R$ is not real. A real representation is obtained by replacing 
$\gamma_R\to i\gamma_R$. This latter basis is used in \cite{blpssw} to show 
that this orientifold has no vector structure.} (For older work on bosonic
and supersymmetric open string orbifolds, see \cite{ps}.)
However, motivated by F-theory orbifolds, it is was found
by Dabholkar and Park \cite{dp} and by Blum and Zaffaroni \cite{bz} that a
\D6 $\IZ_2$ orientifold with vector structure can be obtained. This model 
(called DPBZ in the following) is realized by taking an orientifold group 
$\{1,R,\Omega(-1)^{F_R}R_1,\Omega(-1)^{F_L}R_2\}$, where $R_i$ is the 
inversion of the $i\th$ internal torus, $i=1,2$, and $R=R_1R_2$.
In order to cancel the untwisted tadpoles, two types of $D7$-branes are 
needed: $D7$-branes wrapping the first torus and $D7'$-branes wrapping the 
second torus. The closed string spectrum consists of 17 tensor multiplets 
and 4 hypermultiplets. After applying T-duality in the second torus one expects
to get the Gimon-Polchinski (GP) model with an orientifold group 
$\{1,R,\Omega,\Omega R\}$. But the closed string spectrum of this latter model
is quite different: it has only one tensor multiplet and 20 hypermultiplets. 
It turns out that the T-dual of the DPBZ model has a slightly different 
$\Omega$ projection as compared to the GP model: 
$\Omega_{\rm  DPBZ}=\Omega_{\rm GP}\,T$, where the operator $T$ flips the sign
of the twisted fields at all fixed points. This change of sign in the twisted 
closed string modes is responsible for the enhancement of the number of tensor
multiplets \cite{dp,p}.

Following the argument of \cite{p}, one can see that this change should be 
accompanied by a change in the consistency conditions on the action of the 
orientifold group elements on the Chan-Paton matrices. That consistency 
condition is of the form:
\be  \label{gam_R}
\gamma_R = -\epsilon\,\gamma_\Omega \gamma^\top_R \gamma^{-1}_\Omega,\qquad
{\rm with}\ \epsilon=\left\{\ba{ll} 
                   +1\ &\hbox{for the GP $\Omega$-projection}\\
                   -1  &\hbox{for the DPBZ $\Omega$-projection} \ea\right.
\ee
It follows that in the GP model, $\gamma_R$ has to be antisymmetric (and real)
which implies $\gamma_R^2=-\one$. However, in the DPBZ model, it is possible
to take a representation with vector structure for the $\IZ_2$ group. 
In order to get a supersymmetric model, one should correlate the closed string
sign and the action of $\Omega^2$ on the $77'$ sectors. All these changes 
produce a model with 32 $D7$-branes and 32 $D7'$-branes, with gauge group 
$SO(8)^4 \times SO(8)^4$ and without charged hypermultiplets \cite{dp,bz}. 

The aim of this article is to generalise the orientifold construction of 
\cite{dp,bz} to \D4 and $\Gamma=\IZ_N$. We analyse the models with even $N$,
where the the distinction between the two cases $\mu=\pm1$, with and without 
vector structure, is relevant. All the features of the DPBZ model discussed
above also appear in the four-dimensional models. In particular there is a 
curious interplay between different factors that may appear in type IIB 
orientifold constructions: vector structure $\mu$, the sign $\epsilon$
and the discrete torsion.\footnote{Discrete torsion is not
possible for the models discussed in this article, but only for 
$\bZ_N\times\bZ_M$ orientifolds. These will be analysed in \cite{kr2}.
For a discussion of the various phase factors see \cite{kr1,kr2}.} 
The relation between these signs can be summarized by
\be
\mu_9 = \mu_5 = -\epsilon.
\ee

When trying to construct four-dimensional type IIB orientifold models that
correspond to discrete groups $\Gamma$ which would lead to \D4, \N1 heterotic 
orbifolds, the authors of \cite{z,afiv} found that some of these orientifolds 
are inconsistent. Consider the simplest model that presents this kind of 
inconsistency: the $\IZ_4$ orientifold \cite{afiv}. The problem appears in 
the tadpole cancellation condition of the order-two twisted sector. The fixed 
set in this sector consists of 16 tori. Four of them are located at $\IZ_4$
fixed points in the first two planes. The $\IZ_4$ group acts as a $\IZ_2$ 
inversion on each of these four tori and permutes the remaining 12 tori. 
The $\IZ_4$-invariant set is given by four $T^2/\IZ_2$'s and 6 pairs of tori.
There is a contribution from the Klein bottle to the order-two twisted 
tadpoles at each of the four $\IZ_4$ fixed points in the first two planes
proportional to   
\be  \label{Z4problem}
\frac{16}{V_3} (16+\epsilon\,16)
\ee
where $V_3$ is the volume of the third torus and $\epsilon$ is defined in
(\ref{gam_R}).
This contribution can be interpreted as the sum of the twisted charges of 
the $O9$-plane and the 16 $O5$-planes that wrap the third torus and sit at
16 $\IZ_2$ fixed points in the first two tori. The case without vector 
structure ($\mu=-\epsilon=-1$) has a \nonvan contribution from the 
orientifold planes that can not be cancelled by adding any set of $D$-branes 
because in this sector the twisted charge of the $D$-branes vanishes. Only in 
the non-compact case, when $V_3 \rightarrow \infty$, and in the case with 
vector structure ($\mu=-\epsilon=1$) does this tadpole vanish. The 
impossibility to cancel this tadpole contribution using the standard GP
$\Omega$-projection can also be seen in the effective field theory which
suffers from \nonab gauge anomalies. The same inconsistency was found in
the $\IZ_8$, $\IZ_8'$ and $\IZ_{12}'$ orientifolds without vector structure
\cite{afiv}.

Similarly, Zwart found \cite{z} that the $\IZ_N \times \IZ_M$ orientifolds, 
with $N$ or $M$ a multiple of four, are not consistent. These models have
discrete torsion in the sense of \cite{df,kr1} and use the standard GP
$\Omega$-projection (\ie have no vector structure in $D9$ and three sets of 
$D5$-brane sectors). This inconsistency is not related to uncancelled 
tadpoles but rather to the algebra of the representations. 

However, $\IZ_N$, $N$ even, orientifolds with vector structure can be 
consistently constructed for the discrete groups $\IZ_4$, $\IZ_6$, $\IZ_6'$, 
$\IZ_8$, $\IZ_{12}$ and $\IZ_{12}'$.\footnote{Furthermore, it is 
also possible to construct $\bZ_N \times \bZ_M$ orientifolds, with $N$ or 
$M$ a multiple of four, as will be shown in \cite{kr2}.} Probably the $\IZ_8'$
orientifold also exists, although we were not able to find a consistent 
solution. Orientifolds with vector structure are also possible if one
introduces anti-$D5$-branes \cite{ads,au,aiq,aaads}. However, the models
with antibranes are not supersymmetric.
In section 2 we discuss the construction of the supersymmetric
orientifolds with vector structure: we analyse the closed and open 
string spectrum and the tadpole cancellation conditions. The physical
interpretation of the various signs that appear in this computation leads
us to speculate about the existence of $D$-branes with negative NSNS charge.
In section 3 we 
give an explicit solution of the tadpole conditions for each of the above 
orientifolds. We concentrate on the brane configurations leading to the 
gauge group with maximal rank. This corresponds to locating a maximum number 
of $D5$-branes at the $\IZ_N$ fixed points in such a way that all the 
tadpoles are cancelled. In some cases not all the $D5$-branes can be put at 
the origin but some of them will be needed at other fixed points to cancel 
the Klein bottle contribution. For the $\IZ_8$ and $\IZ_{12}$ orientifold,
it is necessary to put some $D5$-branes at points which are not fixed
under $\IZ_N$. We verify the correspondence between the tadpole cancellation 
conditions and the \nonab gauge anomalies that was expected from the general
analysis of the models without vector structure \cite{abiu,lr}. The Abelian 
gauge anomalies are expected to be cancelled by the four-dimensional 
generalisation of the Green-Schwarz mechanism \cite{iru1,klein}. 
In the two appendices we explain how to obtain the open string spectrum using
orientifold quivers and sketch a derivation of the tadpole cancellation
conditions including all possible phase factors.

\section{Construction of the models}
\label{construction}

We consider compact orientifolds of the form 
$T^6/(\Gamma\times\{\one,\Omega\})$, with $\Gamma=\IZ_N$, $N$ even. The
six-torus is defined as $T^6=\IC^3/\Lambda$, with $\Lambda$ a factorisable
lattice, \ie it is the direct sum of three two-dimensional lattices.%
\footnote{More precisely, we only need that for each 
$g\in\Gamma\backslash\{e\}$
that has fixed planes, the lattice $\Lambda$ can be decomposed in a direct sum
of sublattices $\Lambda=I\oplus J$, such that $I$ is fixed under $g$ and this
decomposition is preserved under $g$ \cite{ek}. The orientifolds corresponding
to $\Gamma=\bZ_8$, $\bZ_8'$, $\bZ_{12}$, $\bZ_{12}'$ only satisfy this weaker
condition.}
The world-sheet symmetry $\Omega$ is of the form
\be \label{OmJT}
\Omega\ =\ \Omega_0\, J\, T,
\ee
where $\Omega_0$ is the world-sheet parity, the operator $J$ exchanges the
$k\th$ and the $(N-k)\th$ twisted sector \cite{p} and $T$ acts as $-\bfm1$ on 
the order-two twisted states. The necessity of the additional operator $T$ is
related to the fact that the models considered in this article have vector
structure. This operator was discussed by the authors of \cite{dp,p} when
analysing a new $\IZ_2$ orientifold in $D=6$.

The action of $\Gamma$ on the coordinates $(z_1,z_2,z_3)$ of $\IC^3$ can be 
characterized by the shift vector $v=(v_1,v_2,v_3)$:
\be  \label{shiftvec}
g_1:\ z_i \longrightarrow e^{2\pi iv_i} z_i,
\ee
where $g_1$ is the generator of $\Gamma$ and $\sum_{i=1}^3v_i=0$ to ensure
\N1 supersymmetry in \D4. Not all possible shifts correspond
to a symmetry of some lattice. Indeed, there is a finite number of
$\IZ_N$ orbifolds (see \eg \cite{ek}). For even $N$, there are only 7 models,
table \ref{ZNorbi}. The shift vectors are chosen such that the order-two
twist $g_1^{N/2}$ fixes the third complex plane. This will require the 
introduction of $D5_3$-branes when constructing the corresponding
orientifolds.

\begin{table}[ht]
\renewcommand{\arraystretch}{1.25}
$$\begin{array}{|c|c||c|c|}
\hline
\Gamma  &v                   &\Gamma &v\\ \hline
\IZ_4   &{1\over4}(1,1,-2)   &       & \\ \hline
\IZ_6   &{1\over6}(1,1,-2)   &\IZ_6'    &{1\over6}(1,-3,2)  \\ \hline
\IZ_8   &{1\over8}(1,3,-4)   &\IZ_8'    &{1\over8}(1,-3,2)  \\ \hline
\IZ_{12}&{1\over12}(1,-5,4)  &\IZ_{12}' &{1\over12}(1,5,-6) \\ \hline
\end{array}$$
\caption{Possible $\bZ_N$ orbifolds with even $N$ and $\cN=1$ in $D=4$.
\label{ZNorbi}}
\end{table}

The $\IZ_N$ orientifolds without vector structure have been discussed in
\cite{ks2,afiv}. There it was found that the models with discrete groups
$\IZ_4$, $\IZ_8$, $\IZ_8'$ and $\IZ_{12}'$ are not consistent because
they have uncancelled tadpoles. We will see that there are solutions to
the tadpole equations for all of the models with vector structure. However,
in the construction of the $\IZ_8'$ orientifold, we encountered a difficulty 
and could not construct a consistent model.

In this section, we sketch the basic steps to construct $\IZ_N$ orientifolds
with vector structure. We explain how to obtain the closed string spectrum, 
the open string spectrum and the tadpole cancellation conditions.

\subsection{Closed string spectrum}
\label{closed}

The closed string spectrum can be obtained from the cohomology of the internal
orbifold space, table \ref{OrbiHodge}. This is explained in \cite{klein,kr1}.
Let us summarize this method and analyse what changes if the 
$\Omega$-projection (\ref{OmJT}) of Dabholkar, Park, Blum and Zaffaroni (DPBZ)
\cite{dp,bz} is taken instead of the standard $\Omega$-projection of Gimon 
and Polchinski (GP) \cite{gp}.


\begin{table}[ht]
\renewcommand{\arraystretch}{1.25}
$$\begin{array}{|c||c|c|c|c|c|c|c|c|c|} \hline
  \Gamma & &k=0 &k=1 &k=2 &k=3 &k=4 &k=5 &k=6 &\rm TOTAL \\
\hline\hline
  \IZ_4 & h^{1,1} & 5 & 16 & 10 & - & - & - & - & 31\\
        & h^{2,1} & 1 & 0  & 6  & - & - & - & - & 7\\
\hline
  \IZ_6 & h^{1,1} & 5 & 3 & 15 & 6 & - & - & - & 29\\
        & h^{2,1} & 0 & 0 & 0  & 5 & - & - & - & 5\\
\hline
  \IZ_6' & h^{1,1} & 3 & 12 & 6 & 8 & - & - & - & 35\\
         & h^{2,1} & 1 & 0  & 3 & 4 & - & - & - & 11\\
\hline
  \IZ_8 & h^{1,1} & 3 & 8 & 3 & 8 & 6 & - & - & 31\\
        & h^{2,1} & 1 & 0 & 1 & 0 & 4 & - & - & 7\\
\hline
  \IZ_8' & h^{1,1} & 3 & 4 & 10 & 4 & 6 & - & - & 27\\
         & h^{2,1} & 0 & 0 & 0  & 0 & 3 & - & - & 3\\
\hline
  \IZ_{12} & h^{1,1} & 3 & 3 & 3 & 2 & 9 & 3 & 4 & 29\\
           & h^{2,1} & 0 & 0 & 0 & 1 & 0 & 0 & 3 & 5\\
\hline
  \IZ_{12}' & h^{1,1} & 3 & 4 & 1 & 8 & 3 & 4 & 4 & 31\\
            & h^{2,1} & 1 & 0 & 0 & 0 & 2 & 0 & 2 & 7\\
\hline
\end{array}$$
\caption{Contribution from each sector to the orbifold Hodge numbers ($k=0$ is
the untwisted sector). The sectors corresponding to twists $k>N/2$ are not 
displayed. In the last column all the contributions are summed up (including 
untwisted and $k>N/2$ sectors). \label{OrbiHodge}}
\end{table}

We can split the sectors into three different types:

(i) The untwisted sector ($k=0$), it is invariant under $J$ and $T$.
The bosonic fields in $D=4$ are found contracting the Lorentz indices
of the $\Omega$-even 10D fields $g_{\mu\nu}$, $\phi$, $C^{(2)}_{\mu\nu}$ with
the harmonic forms corresponding to $h^{0,0}$, $h^{3,0}$, $h^{1,1}_{\rm untw}$,
$h^{2,1}_{\rm untw}$.

(ii) The order-two sector ($k=N/2$), it is invariant under $J$ but acquires
an extra minus sign under the action of $T$. In this sector, one has to 
contract the $\Omega$-odd 10D fields $B_{\mu\nu}$, $C^{(4)}_{\mu\nu\rho\sigma}$
with $h^{1,1}_{N/2}$, $h^{2,1}_{N/2}$. Only in this sector does the spectrum
differ from the one obtained when using the standard GP $\Omega$-projection.

(iii) The remaining sectors. To get the fields in $D=4$, one forms linear
combinations of the harmonic forms that belong to the $k\th$ and $(N-k)\th$
twisted sector. The $J$-even forms are contracted with the $\Omega$-even 10D 
fields and the $J$-odd forms are contracted with the $\Omega$-odd 10D fields.

The spectrum fits into $\cN=1$ supermultiplets. In total, one finds:

(i) the gravity multiplet, a linear multiplet, $(h^{1,1}+h^{2,1})_{\rm untw}$ 
chiral multiplets.

(ii) $h^{1,1}_{N/2}$ linear multiplets, $h^{2,1}_{N/2}$ vector multiplets.

(iii) for each $0<k<N/2$: $h^{1,1}_k$ linear multiplets, $h^{2,1}_k$ vector 
multiplets; if the $k\th$ sector has fixed planes, then there are additional
$(h^{1,1}_k+h^{2,1}_k)$ chiral multiplets.

\subsection{Open string spectrum}
\label{open}

There are 32 $D9$-branes and 32 $D5_3$-branes, the index 3 indicating that 
the 5-branes fill the four non-compact directions and the third complex 
plane. This is a consequence of the untwisted tadpole cancellation 
conditions to be discussed below.

The action of $\Gamma$ on the Chan-Paton indices of the open strings is
described by a (projective) representation $\gamma^{(p)}$ that associates a 
$(32\times32)$-matrix $\gamma_{g,p}$ to each element $g$ of $\Gamma$, where
$p=9,5$ denotes the type of the $D$-brane the open string ends on.
\bea  \label{def_gamma}
\gamma^{(p)}:\ \Gamma &\longrightarrow &GL(32,\IC) \\
                    g &\longmapsto     &\gamma_{g,p} \nonumber
\eea
Because of the orientifold projection, this representation must be real or
pseudo-real. In general, $\gamma^{(p)}$ can be decomposed in irreducible 
blocks of real ($R^r$), pseudo-real ($R^p$) and complex ($R^c$) representations
\cite{bi}:
\be  \label{gamma_rep}
\gamma^{(p)}=\left(\bigoplus_{l_1}n^r_{l_1}R^r_{l_1}\right)\oplus
             \left(\bigoplus_{l_2}n^p_{l_2}R^p_{l_2}\right)\oplus 
         \left(\bigoplus_{l_3}n^c_{l_3}(R^c_{l_3}\oplus\bar R^c_{l_3})\right).
\ee
In this expression, the notation $n_lR_l$ is short for $R_l\otimes\one_{n_l}$,
\ie $n_l$ is the number of copies of the irreducible representation (irrep) 
$R_l$ in $\gamma^{(p)}$ \cite{bi}. 

Let us first consider the 99 and $5_35_3$ sectors. The projection on invariant
states of the Chan-Paton matrices $\lambda^{(0)}$ that correspond to gauge
bosons in $D=4$ imposes the constraints
\be  \label{gauge_proj}
\lambda^{(0)}=\gamma_{g_1,p}\lambda^{(0)}\gamma_{g_1,p}^{-1},\qquad
\lambda^{(0)}=-\alpha_p\,\gamma_{\Omega,p}{\lambda^{(0)}}^\top
                         \gamma_{\Omega,p}^{-1},
\ee
where $g_1$ is the generator of $\Gamma$. The sign $\alpha_p$ is related
to the $D$-brane charges. In the models we consider, $\alpha_9=1$ and
$\alpha_5=-1$. The sign $\alpha\equiv\alpha_9\alpha_5$ also appears in the
$95_3$ cylinder contribution to the tadpoles. For a further discussion, see
section \ref{interpretation}.
The constraints (\ref{gauge_proj}) are easily solved. One finds \cite{bi} 
that the gauge group on the $Dp$-branes is 
\be  \label{Gp_sym}
G_{(p)}\ =\ \prod_{l_1}SO(n^r_{l_1})\times\prod_{l_2}USp(n^p_{l_2})\times
            \prod_{l_3}U(n^c_{l_3})
\ee
if $\gamma_{\Omega,p}$ is symmetric and $\alpha_p=1$ or if
if $\gamma_{\Omega,p}$ is antisymmetric and $\alpha_p=-1$ and
\be  \label{Gp_asym}
G_{(p)}\ =\ \prod_{l_1}USp(n^r_{l_1})\times\prod_{l_2}SO(n^p_{l_2})\times
            \prod_{l_3}U(n^c_{l_3})
\ee
if $\gamma_{\Omega,p}$ is antisymmetric and $\alpha_p=1$ or if
if $\gamma_{\Omega,p}$ is symmetric and $\alpha_p=-1$.

In our case, $\Gamma=\IZ_N$. There are $N$ irreps, all of them one-dimensional:
\be  \label{ZN_irreps}
R_0(g_1)=1,\quad R_1(g_1)=e^{2\pi i/N}, \quad\ldots\quad, \quad 
R_{N-1}(g_1)=e^{2\pi i(N-1)/N}.
\ee
These irreps have vector structure because $(R_l)^N=1\ \forall\,l$.
For even $N$, $R_0$ and $R_{N/2}$ are real and the remaining irreps can be
organized in pairs of conjugate representations: $R_l=\bar R_{N-l}$.
Furthermore, we will see below that $\gamma_{\Omega,9}$ is symmetric,
$\alpha_9=1$ and $\gamma_{\Omega,5}$ is antisymmetric, $\alpha_5=-1$. 
Thus the gauge group is
\be  \label{gauge_gr}
G_{(9)}\ =\ G_{(5)}\ =\ SO(n_0)\times SO(n_{N/2})\times
                        \prod_{l=1}^{N/2-1}U(n_l).
\ee

The matter fields corresponding to the $i\th$ complex plane are obtained from
the projections
\bea  \label{matter_proj}
\lambda^{(i)}=e^{2\pi iv_i}\gamma_{g_1,p}\lambda^{(i)}\gamma_{g_1,p}^{-1},
&\quad
&\lambda^{(i)}=\alpha_p\,R_{\Omega}^{(i)}\gamma_{\Omega,p}{\lambda^{(i)}}^\top
              \gamma_{\Omega,p}^{-1}, \\
&\quad
&{\rm with}\ R_{\Omega}^{(i)}=\left\{\begin{array}{ll}
                                -1\ &{\rm if\ }p=9\ {\rm or\ }i=3\\
                                +1\ &{\rm if\ }p=5\ {\rm and\ }i\neq3
                                    \end{array}\right.. \nonumber
\eea
These equations are solved in two steps. First, one draws the quiver diagram
of the corresponding $\IZ_N$ orbifold \cite{dm}. This solves the first 
condition of (\ref{matter_proj}). Second, the orientifold projection is 
performed on this quiver diagram, as explained in appendix \ref{app_quiver}. 
The spectrum can be directly read off from the resulting orientifold quiver.

Let us now consider the $95_3$ sector. The projection of the open strings
$\lambda^{(95)}$, stretching from 9-branes to 5-branes, on $\Gamma$-invariant
states reads (see \eg \cite{afiv}):
\be  \label{mixed_proj}
\lambda^{(95)}=e^{-\alpha\,\pi iv_3}\gamma_{g_1,9}\lambda^{(95)}
               \gamma_{g_1,5}^{-1}
\ee
Here, we used $\sum_{i=1}^3v_i=0$ to transform $e^{2\pi i(-1/2)(v_1+v_2)}=
e^{\pi iv_3}$. The additional sign $\alpha\equiv\alpha_9\alpha_5$ in the 
exponent is related to the charge difference between the $D9$-branes and
the $D5_3$-branes.\footnote{The relation between $\alpha$ and the $D$-brane
charges is explained in section \ref{interpretation}.} In orientifold models
containing antibranes, a similar sign appears in the R sector \cite{au},
whereas the fields of the NS sector satisfy a different projection equation.
In contrast to this, our models are supersymmetric and therefore, the
projections in the R and NS sectors coincide. $\Omega$ relates the $95_3$ 
sector with the $5_39$ sector and does not impose extra conditions on 
$\lambda^{(95)}$. Again, the spectrum is easiest obtained using quiver 
theory, see appendix \ref{app_quiver}.

\subsection{Tadpoles}
\label{tadpoles}

In this paragraph we give the tadpole cancellation conditions keeping 
track of all possible signs. This enables us to see how the tadpole 
equations of the models with DPBZ projection differ from those with GP 
projection. The derivation of these equations is sketched in appendix 
\ref{app_tad}.

It is convenient to label the elements of $\IZ_N$ by the integer 
$k=0,\ldots,N-1$, \ie $k=0$ denotes the neutral element and $k=1$ the
generator of $\IZ_N$. We define $s_i=\sin(\pi kv_i)$, $c_i=\cos(\pi kv_i)$ 
and $\tilde s_i=\sin(2\pi kv_i)$.
Let us comment on the various signs appearing in the tadpole contributions.
\begin{itemize}
\item In the cylinder amplitude, $\alpha$ weights the $95_3$ sector relative
      to the 99 and $5_35_3$ sectors. It is related to the signs appearing
      in (\ref{gauge_proj}) and (\ref{matter_proj}) by 
      $\alpha=\alpha_9\alpha_5$. One has $\alpha=1$ for the standard
      orientifold constructions \cite{gp,z,afiv}. For the \nonsusy
      orientifolds containing antibranes, $\alpha=-1$ in the RR sector
      but $\alpha=1$ in the NSNS sector \cite{ads,au,aaads}. In our case, 
      $\alpha=-1$ in the RR sector and in the NSNS sector.

\item In the Klein bottle amplitude, $\epsilon$ is related to the choice
      between the standard and alternative $\Omega$-projection. One has
      $\epsilon = +1$ for GP and $\epsilon = -1$ for DPBZ.\footnote{The
      same sign $\epsilon$ also appears in the \nonsusy models of 
      \cite{aaads}.}

\item In the M\"obius strip, $\mu_p$ and $c_p$ are defined by 
      $(\gamma_{1,p})^N=\mu_p\,\one$ and 
      $\gamma_{\Omega,p}^\top=c_p\,\gamma_{\Omega,p}$.
      Models with $\mu_p=+1$ ($-1$) are said to have (no) vector structure.
\end{itemize}
A more precise definition of these signs is given in appendix \ref{app_tad}.
For a physical interpretation, see section \ref{interpretation}.

Factorisation of the twisted tadpoles requires that 
\be  \label{constr_fac}
\epsilon=-\mu_9=-\mu_5.
\ee
This relates the sign $\epsilon$ of the twisted states and the vector 
structure. One can easily check using the orientifold relations and the 
unitarity of the matrices that this condition is equivalent to the consistency 
condition (\ref{gam_R}) mentioned by Polchinski in \cite{p}.

The untwisted tadpoles can only be cancelled if
\be  \label{alpha_epsilon}
\alpha=\epsilon.
\ee
According to the interpretation of $\alpha$ and $\epsilon$ given in section
\ref{interpretation}, this is the statement that the $D$-brane charges must
be opposite to the $O$-plane charges.

Furthermore, it is easy to see \cite{gp} that the action of $\Omega^2$ on
the oscillator part of strings stretching from 9-branes to 5-branes is
related to $c_9$ and $c_5$ by 
\be  \label{Om95}
\Omega^2|_{95_3}\ =\ c_9c_5\ =\ -1.
\ee           
The last equality follows from the arguments given in \cite{gp}. In special
models with $95_3$ matter only at half-integer mass levels, one can have
$\Omega^2|_{95_3}=1$, as shown in \cite{dp}. However, all of our models 
have massless matter in the $95_3$ sector. Thus, we need 
$\Omega^2|_{95_3}=-1$.

A supersymmetric solution does only exist if
\be  \label{constr_untw}
c_9=1.
\ee

We are interested in the supersymmetric models with vector structure, \ie
\be  \label{all_signs} 
c_9=\mu_9=\mu_5=1,\qquad \Omega^2|_{95_3}=c_5=\epsilon=\alpha=-1.
\ee

The tadpole cancellation conditions for supersymmetric models are:

a) untwisted sector
\be  \label{tad_untw}
  \Tr\gamma_{0,9} = 32, \qquad
  \Tr\gamma_{0,5} = 32\,c_5\,\mu_5
\ee

b) twisted sectors without fixed tori, \ie $kv_i\ne0\ {\rm mod\ }\IZ$:
\begin{itemize}
\item odd $k$:
\be  \label{tad_oddk}  
   \Tr\gamma_{k,9} + 4\,\alpha\,s_1 s_2 \Tr\gamma_{k,5}=0
\ee

\item even $k=2k'$:
\be  \label{tad_evenk}  
  \Tr\gamma_{2k',9} + 4\,\alpha\,\tilde s_1\tilde s_2 \Tr\gamma_{2k',5}
     = 32\,(c_1c_2c_3 + \epsilon\,s_1s_2c_3),
\ee
where $s_i$, $c_i$, $\tilde s_i$ are evaluated with the argument $k'$.
\end{itemize}

c) twisted sectors with fixed tori, \ie $kv_i=0\ {\rm mod\ }\IZ$:
\begin{itemize}
\item odd $k$:
  \begin{itemize}
  \item[-] $i=3$:
  \be  \label{tad_fixodd}
     \Tr\gamma_{k,9} + 4\,\alpha\,s_1 s_2 \Tr\gamma_{k,5}=0
  \ee

  \item[-] $i\ne3$: \qquad never happens
  \end{itemize}

\item even $k=2k'$, with $k'v_i=0$:
  \begin{itemize}
  \item[-] $i=3$:
  \be  \label{tad_fixeven1}  
     \Tr\gamma_{2k',9} + 4\,\alpha\,\tilde s_1\tilde s_2\Tr\gamma_{2k',5}
         = 32\,(c_1c_2 + \epsilon\,s_1s_2)
  \ee

  \item[-] $i\ne3$:
  \be  \label{tad_fixeven2}  
     \Tr\gamma_{2k',9} = - 32\,c_3^2,\qquad
     \Tr\gamma_{2k',5} = -  8\,c_5\,\mu_5
  \ee
  \end{itemize}

\item even $k=2k'$, with $k'v_i=\pm\half$:
  \begin{itemize}
  \item[-] $i=3$:
  \be  \label{tad_fixeven3}  
     \Tr\gamma_{2k',9} + 4\,\alpha\,\tilde s_1\tilde s_2\Tr\gamma_{2k',5} 
           = 0,\qquad
     \epsilon = -1
  \ee

  \item[-] $i\ne3$:
  \be  \label{tad_fixeven4}  
     \Tr\gamma_{2k',9} = \mp 32\,\mu_9\,c_3^2,\qquad
     \Tr\gamma_{2k',5} = \mp  8\,c_5
  \ee
  \end{itemize}
\end{itemize}

\noindent Note that the second condition in (\ref{tad_fixeven3}) can only
be satisfied in models with vector structure, $\mu_9=\mu_5=1$. This tadpole
arises whenever $v_3=\pm\frac{1}{2p}$ for some integer $p$, which is only
possible for $N$ a multiple of 4 because we want the third torus to be
fixed under $k=N/2$. A short look at the shift vectors of table \ref{ZNorbi}
shows that the orientifolds with discrete groups $\IZ_4$, $\IZ_8$, $\IZ_8'$
and $\IZ_{12}'$ have this property. Therefore, they are only consistent with
vector structure.

The $D5_3$-branes may be distributed over different points in the first two
internal tori. Of course the tadpole equations depend on the location of
the 5-branes and not all configurations are consistent. Three different
types of points have to be distinguished:
\begin{itemize}
\item $\IZ_N$ fixed points, like the origin,

\item $\IZ_M$ fixed points (with $M$ a divisor of $N$) which are not fixed 
      under $\IZ_N$,

\item Points in the bulk.
\end{itemize}
The Klein bottle contribution to the tadpoles of the $k$-twisted sector
consists of an untwisted part $\cK_0(k)$ and a twisted part $\cK_1(k)$.
The former gives the term proportional to $c_1c_2c_3$ in (\ref{tad_evenk}),
the latter the term proportional to $\epsilon\,s_1s_2c_3$. (Note that these
terms are combined with the cylinder contribution to the $2k$-twisted sector.)
At each point of the internal space, the contribution $\cK_0(k)$ is only 
present if this point is fixed under $k$ and the contribution $\cK_1(k)$ is 
only present if this point is fixed under $k+N/2$. Thus, the above tadpole 
cancellation conditions are strictly valid only at the $\IZ_N$ fixed points. 
At $\IZ_M$ fixed points which are not fixed under $\IZ_N$, these conditions 
have to be modified accordingly.

One has to analyse the tadpoles at all the fixed points in each twisted sector.
At each fixed point all the $D9$-branes contribute but only those $D5_3$-branes
that are located at this point. The Klein bottle contribution to this fixed
point is determined as explained in the preceding paragraph.
We will see how this works in the examples below. In the some of the models 
it is not possible to put all the $D5_3$-branes at the origin. 

The twisted tadpole conditions from sectors that do not fix the third torus
are in one-to-one correspondence to the conditions arising from the 
requirement of absence of \nonab gauge anomalies. This correspondence has
been studied for orientifolds with GP projection in \cite{abiu}.

\subsection{A physical interpretation of the signs $\epsilon$ and $\alpha$}
\label{interpretation}
It is instructive to analyse the relation between the signs $\epsilon$ and
$\alpha$ and the charges of the $D$-branes and $O$-planes. These signs
are defined in the second paragraph of the previous subsection and in
appendix \ref{app_tad}.

The \D6 $\IZ_2$ orientifold of Gimon and Polchinski \cite{gp} (which was
first constructed by Bianchi and Sagnotti \cite{bs}) contains
one $O9^-$-plane (of RR charge $-32$ and NSNS charge $-32$) and 16 
$O5^-$-planes (each of RR charge $-2$ and NSNS charge $-2$) extended in 
the 6 space-time dimensions and located at the 16 fixed points of $\Omega R$ 
in the compact dimensions. Tadpole cancellation requires the introduction of 
32 $D9$-branes and 32 $D5$-branes (each of RR charge $+1$ and NSNS charge 
$+1$). This model is supersymmetric. Indeed, the RR tadpoles and the
NSNS tadpoles vanish both. Looking at the specific form of the Klein bottle
and cylinder amplitude, we find that $\epsilon=\alpha=1$ in the GP model.

Antoniadis, Dudas and Sagnotti \cite{ads} and Aldazabal and Uranga \cite{au}
construct a different \D6 $\IZ_2$ orientifold by introducing a sign 
$\epsilon=-1$ in the twisted Klein bottle contribution. This corresponds
to replacing the 16 $O5^-$-planes of the GP model by 16 $O5^+$-planes (each
of RR charge $+2$ and NSNS charge $+2$). Thus, $\epsilon$ is the relative
sign between the charges of the $O9$-planes and the $O5$-planes \cite{ads}.
To cancel the RR tadpoles, the authors of \cite{ads,au} introduce 32 
anti-$D5$-branes (each of RR charge $-1$ and NSNS charge $+1$). This leads
to a sign flip of the coefficient $\alpha$ in the cylinder contribution to
the RR tadpoles (\eg in eq.\ (\ref{tad_evenk})). The corresponding 
contribution to the NSNS tadpoles remains unchanged. Thus, one has 
$\alpha^{R}=-1$, $\alpha^{NS}=+1$. We deduce that $\alpha$ is the relative
sign between the charges of $D9$-branes and $D5$-branes. This model
is not supersymmetric. Indeed, there is a clear asymmetry between the RR
tadpoles and the NSNS tadpoles. The former vanish whereas the latter lead
to a \nonvan potential for the dilaton. 

The orientifold of Dabholkar and Park \cite{dp} and of Blum and Zaffaroni
\cite{bz} is yet another \D6 $\IZ_2$ model. They also introduce a sign
$\epsilon=-1$ in the twisted Klein bottle contribution. This is a consequence
of the operator $T$ in the DPBZ $\Omega$-projection (\ref{OmJT}). 
Geometrically, this means that their model contains 16 $O5^+$-planes. This
leads to a puzzle: On the one hand the DPBZ model is clearly supersymmetric
(it has an equivalent description as a supersymmetric F-theory vacuum),
on the other hand there is no obvious object that could cancel the NSNS
charge of the $O5^+$-planes. The authors of \cite{dp} noticed that the
sign in front of the twisted 95 cylinder amplitude is changed. One has 
$\alpha=-1$ for this model. Unlike the case of the previous paragraph, 
this sign flip affects the RR tadpoles as well as the NSNS tadpoles. 
Assuming that the interpretation of $\alpha$ as the charge difference 
between $D9$-branes and $D5$-branes is also correct in the case of NSNS
charges, this leads us to propose the following solution 
to the above puzzle: The $D5$-branes present in the DPBZ model have the 
peculiar property of having RR charge $-1$ and NSNS charge $-1$. They might 
be called $D5^-$-branes because their relation to the standard $D5$-branes 
is the same as the relation of the $O5^-$-planes to the $O5^+$-planes.

The sign flip in the NSNS charge of the $D5$-branes has intriguing
consequences and deserves some further comments.\footnote{We would like 
to thank A.\ Sagnotti, C.\ Angelantonj and G.\ d'Appollonio for useful 
discussions about the meaning, interpretation and consistency of this 
sign flip.} We have argued that the sign $\alpha^{NS}$ can be interpreted 
as the relation between the NSNS charges of the  $D9$-branes and the 
$D5$-branes. Choosing a positive charge for the $D9$-branes, we can 
understand the $D5^-$-branes as $D5$-branes with NSNS and RR charges 
reversed, $\alpha^{NS}=\alpha^{R}=-1$. $D5^-$-branes are $D5$-branes 
with negative tension and RR charge inside $D9$-branes with positive 
tension. The sign $\alpha^{R}=-1$ in models with antibranes is well 
understood. In the open string channel $\alpha^{NS}=1$ and $\alpha^{R}=-1$ 
lead to the opposite GSO projection in the $95$ sector resulting in a
\nonsusy spectrum. In the case we are considering here, 
$\alpha^{NS}=\alpha^{R}=-1$, the GSO projection is the usual supersymmetric 
one but a global minus sign appears in front of the one-loop open string 
partition function in the $D9D5^-$ sector. In the usual ($\alpha^{NS}=1$) 
one-loop partition function of the $95$ sector, the contribution of the
NS sector (Lorentz scalars) is weighted with a plus sign and the contribution
of the R sector (Lorentz spinors) with a minus sign. This reflects the fact
that spinors obey the Fermi statistics. The minus sign for fermion loops is 
a standard result of quantum field theory. In the models we are dealing with,
this relation is reversed due to the global sign change. The NS sector is 
weighted as fermionic and the R sector as bosonic. This seems to contradict
the spin-statistics theorem. It is not clear to us if there is another 
interpretation of this minus sign.

The question whether this sign flip is possible or not might be answered by 
considering higher-loop amplitudes. For closed string theories, the authors
of \cite{abk} found a relation between the spin-statistics correspondence 
in the one-loop partition function and higher-loop modular invariance.
This has been generalised to open string theories \cite{bs2}. The idea is 
the following: one can consider the double cover of an open string diagram 
(\eg the torus for the cylinder or the Klein bottle), determine the subgroup 
of the modular group compatible with the boundaries and apply that group to 
several factorisation limits. This group mixes different one-loop amplitudes 
leading to some constraints on the allowed signs. For theories containing
only $D9$-branes, one finds that the usual spin-statistics correspondence
in the one-loop partition function is implied by higher-loop modular 
invariance. In our case, one should consider diagrams containing different 
types of boundaries, corresponding to $D9$- and $D5$-branes. It is not clear 
to us whether the argument of \cite{bs2} can be generalised to this case
nor what would be the modular group compatible with types of boundaries. 
In particular, the signs $\alpha$ can only be determined if different 
boundaries can be permuted.

Summarizing, $\alpha^{NS}=-1$ in the closed string channel leads to negative 
tension $D5$-branes inside positive tension $D9$-branes and  in the open
string channel it causes a global minus sign in the $95$ one-loop partition 
function. The correct interpretation of this sign and the relation to 
higher-loop amplitudes still need to be understood.

The orientifold models that we present in this article are the direct
generalisation of the DPBZ model to \D4 and orbifold groups $\IZ_N$.%
\footnote{However, in contrast to \cite{dp,bz}, our models have
$\Omega|_{95}=-1$.}
All these models have $\epsilon=\alpha=-1$. Therefore one might think
of our models as containing 16 $O5^+$-planes and 32 $D5^-$-branes.

\section{Description of the models}
\label{description}

All the models contain 32 $D9$-branes and 32 $D5_3$-branes (wrapping the third
internal torus). In general there are two different orientifolds for each
orbifold group $\Gamma$.

(i) The orientifold without vector structure, \ie $\mu_9=\mu_5=-1$. For
consistency, we need to have $\gamma_{\Omega,9}$ symmetric and
$\gamma_{\Omega,5}$ antisymmetric. This is the usual GP projection,
with $\epsilon=\alpha=1$.

(ii) The orientifold with vector structure, \ie $\mu_9=\mu_5=1$. This is the 
alternative projection of DPBZ, with $\epsilon=\alpha=-1$.
In contrast to the original DPBZ orientifold, our models contain massless
matter in the $95_3$ sector. Therefore, the argument of \cite{gp} applies
and we need to have $\gamma_{\Omega,9}$ symmetric and $\gamma_{\Omega,5}$ 
antisymmetric. 

It turns out that some of the models of type (i) are not consistent because
they have uncancelled tadpoles. This inconsistency can also be seen as the 
impossibility to find a brane configuration that leads to a gauge anomaly free
spectrum. (The equivalence of these two point of views has been studied in
\cite{abiu}.) However, there exists at least one solution to the tadpole
equations for each of the models of type (ii).\footnote{The case of the
$\bZ_8'$ orientifold has to be revisited, see below.} In this section, we 
give the complete spectrum of all these models.

\subsection{$\bZ_4$, $v={1\over4}(1,1,-2)$}
\label{Z4}

It is possible to put all the $D5_3$-branes at the origin of the first 
two tori. The tadpole cancellation conditions for this configuration are:
\bea
k=0: &\ &\Tr(\gamma_{0,9}) = \Tr(\gamma_{0,5}) = 32  \nonumber\\
k=1: &  &\Tr(\gamma_{1,9}) -2\, \Tr(\gamma_{1,5}) = 0,\qquad
         \Tr(\gamma_{1,9}) = 0 \\
k=2: &  &\Tr(\gamma_{2,9}) -4\, \Tr(\gamma_{2,5}) = 0,\qquad
         \Tr(\gamma_{2,9}) = 0 \nonumber
\eea
The first condition for the sectors $k=1,2$ corresponds to the fixed
point at the origin; the second condition corresponds to the remaining
fixed points, where no $D5_3$-branes are present. All these conditions 
can be satisfied simultaneously. The solution is unique and leads to the
spectrum displayed in table \ref{Z4spec}. The matter representations are
denoted by their Young tableaux, the indices correspond to the number of
the gauge group factor under which these fields transform.

\begin{table}[h]
\renewcommand{\arraystretch}{1.25}
$$\begin{array}{|c|c|}
\hline
\multicolumn{2}{|c|}{\IZ_4,\ v=\frac14(1,1,-2),\ \mu=1}\\ \hline\hline
\multicolumn{2}{|c|}{\hbox{open string spectrum}}\\ \hline
\rm sector &\hbox{gauge group / matter fields}\\ \hline
99    &SO(8)_1\times U(8)_2\times SO(8)_3\\
      &2\,(\Yfun_1,\Yfunb_2),\ 2\,(\Yfun_2,\Yfun_3),\ 
       (\Yfun_1,\Yfun_3),\ \Yasym_2,\ \Yasymb_2\\ \hline
55    &\hbox{identical to 99}\\ \hline
95    &(\Yfun_1,\Yfun_2),\ (\Yfun_2,\Yfun_1),\ (\Yfunb_2,\Yfun_3)\, 
       (\Yfun_3,\Yfunb_2)\\ \hline\hline
\multicolumn{2}{|c|}{\hbox{closed string spectrum}}\\ \hline
\rm sector  &\hbox{$\cN=1$ multiplets}\\ \hline
\rm untw.   & \hbox{gravity, 1 lin., 6 chir.} \\ \hline
k=N/2       & \hbox{10 lin., 6 vec.} \\ \hline
0<k<N/2     & \hbox{16 lin.} \\ \hline
\end{array}$$
\caption{Spectrum of the $\bZ_4$ orientifold with vector structure.
         \label{Z4spec}}
\end{table}

This model is selfdual under T-duality in the first and second torus.

Following the argument of \cite{abiu}, one can see the relation between the 
\nonab anomalies and the tadpoles of the $k=1$ sector. (The sectors $k=0,2$ 
fix the third torus and give additional constraints.) The gauge group of the
$\IZ_4$ orientifold with vector structure has the general form 
$SO(n_1)\times U(n_2)\times SO(n_3)$. The conditions on the numbers 
$n_1,n_2,n_3$ that are necessary for an anomaly free spectrum can be rewritten
in terms of traces of $\gamma$ matrices: 
$-2\,\Tr(\gamma_{1,9})+\Tr(\gamma_{1,5})=0$ (for the 99 gauge group) and
$\Tr(\gamma_{1,9})-2\,\Tr(\gamma_{1,5})=0$ (for the 55 gauge group). These two
equations are equivalent to the two tadpole equations of the $k=1$ sector.
 
Notice the similarity of this model with the \nonsusy $\IZ_4$ orientifold of
\cite{aaads}. The difference is that the authors of \cite{aaads} use the
standard $\Omega$ action of GP, $\Omega^2|_{95_3}=-1$, which requires 32 
anti-$D5_3$-branes to cancel the untwisted tadpoles. For consistency, 
$\gamma_{\Omega,5}$ has to be antisymmetric. According to the general 
formula (\ref{Gp_asym}) this gives symplectic instead of orthogonal gauge
group factors. Therefore, the gauge group of the 55 sector of their model 
is $USp(8)\times U(8)\times USp(8)$.

\subsection{$\bZ_6$, $v={1\over6}(1,1,-2)$}
\label{Z6}

It is possible to put all the $D5_3$-branes at the origin of the first 
two tori. The tadpole cancellation conditions for this configuration are:
\bea
k=0: &\ &\Tr(\gamma_{0,9}) = \Tr(\gamma_{0,5}) = 32 \nonumber\\
k=1: &  &\Tr(\gamma_{1,9})-\Tr(\gamma_{1,5}) = 0 \\
k=2: &  &\Tr(\gamma_{2,9})-3\,\Tr(\gamma_{2,5}) = 8,\qquad
         \Tr(\gamma_{2,9}) = -4 \nonumber\\
k=3: &  &\Tr(\gamma_{3,9})-4\,\Tr(\gamma_{3,5}) = 0,\qquad
         \Tr(\gamma_{3,9}) = 0 \nonumber
\eea
Again, the first condition for each twisted sector corresponds to the 
fixed point at the origin; the second condition corresponds to the 
remaining fixed points, where no $D5_3$-branes are present. In the $k=2$
sector, there are 9 fixed points in the first two tori. The untwisted
Klein bottle contribution $\cK_0(1)$ is only present at the origin (this
is the only fixed point under $k'=k/2=1$). The twisted Klein bottle
contribution $\cK_1(1)$ is present at all of the 9 fixed points (all of
them are fixed under $k'=k/2+N/2=4$). This explains the right hand side
of the $k=2$ tadpole conditions. There are five solutions to these equations.
The gauge group and the matter are shown in table \ref{Z6spec}.

\begin{table}[h]
\renewcommand{\arraystretch}{1.25}
$$\begin{array}{|c|c|}
\hline
\multicolumn{2}{|c|}{\IZ_6,\ v=\frac16(1,1,-2),\ \mu=1}\\ \hline\hline
\multicolumn{2}{|c|}{\hbox{open string spectrum}}\\ \hline
\rm sector &\hbox{gauge group / matter fields}\\ \hline
99    &SO(8-2n)_1\times U(8-n)_2\times U(4+n)_3\times SO(2n)_4\\
      &2\,(\Yfun_1,\Yfunb_2),\ 2\,(\Yfun_2,\Yfunb_3),\  2\,(\Yfun_3,\Yfun_4),\ 
       (\Yfun_1,\Yfun_3),\ (\Yfunb_2,\Yfun_4),\ \Yasym_2,\ \Yasymb_3 \\ \hline
55    &\hbox{identical to 99}\\ \hline
95    &(\Yfun_1,\Yfun_2),\ (\Yfun_2,\Yfun_1),\ (\Yfunb_2,\Yfun_3),\ 
       (\Yfun_3,\Yfunb_2),\ (\Yfunb_3,\Yfun_4),\ (\Yfun_4,\Yfunb_3)\\ 
\hline\hline
\multicolumn{2}{|c|}{\hbox{closed string spectrum}}\\ \hline
\rm sector  &\hbox{$\cN=1$ multiplets}\\ \hline
\rm untw.   & \hbox{gravity, 1 lin., 5 chir.} \\ \hline
k=N/2       & \hbox{6 lin., 5 vec.} \\ \hline
0<k<N/2     & \hbox{18 lin.} \\ \hline
\end{array}$$
\caption{Spectrum of the $\bZ_6$ orientifold with vector structure.
         The parameter $n$ can take the five values $n=0,\ldots,4$.
         \label{Z6spec}}
\end{table}

All the five models are selfdual under T-duality in the first two tori.
One can verify that the conditions for absence of \nonab gauge anomalies 
coincide with the tadpole conditions for the sectors $k=1,2$.

For completeness and for comparison, we also give the spectrum of the
$\IZ_6$ orientifold without vector structure, table \ref{Z6spec_novs}. 
This orientifold has been analysed by the authors of \cite{ks2,afiv,abiu}.
There are 13 consistent models with all the $D5_3$-branes at the origin. 
One of them is selfdual under T-duality in the first two tori. The remaining 
12 are organized in pairs of T-dual models.

\begin{table}[h]
\renewcommand{\arraystretch}{1.25}
$$\begin{array}{|c|c|}
\hline
\multicolumn{2}{|c|}{\IZ_6,\ v=\frac16(1,1,-2),\ \mu=-1}\\ \hline\hline
\multicolumn{2}{|c|}{\hbox{open string spectrum}}\\ \hline
\rm sector &\hbox{gauge group / matter fields}\\ \hline
99    &U(n)_1\times U(4)_2\times U(12-n)_3\\
      &2\,(\Yfun_1,\Yfunb_2),\ 2\,(\Yfun_2,\Yfunb_3),\ 
       (\Yfun_1,\Yfun_2),\ (\Yfunb_2,\Yfunb_3),\ (\Yfun_3,\Yfunb_1),\ 
       2\,\Yasymb_1,\ 2\,\Yasym_3 \\ \hline
55    &U(12-n)\times U(4)\times U(n)\\
      &\hbox{matter identical to 99}\\ \hline
95    &(\Yfunb_1,\Yfunb_1),\ (\Yfun_1,\Yfunb_2),\ (\Yfunb_2,\Yfun_1),\ 
       (\Yfun_2,\Yfunb_3),\ (\Yfunb_3,\Yfun_2),\ (\Yfun_3,\Yfun_3)\\ 
\hline\hline
\multicolumn{2}{|c|}{\hbox{closed string spectrum}}\\ \hline
\rm sector  &\hbox{$\cN=1$ multiplets}\\ \hline
\rm untw.   & \hbox{gravity, 1 lin., 5 chir.} \\ \hline
k=N/2       & \hbox{11 chir.} \\ \hline
0<k<N/2     & \hbox{18 lin.} \\ \hline
\end{array}$$
\caption{Spectrum of the $\bZ_6$ orientifold without vector structure.
         The parameter $n$ can take the 13 values $n=0,\ldots,12$.
         \label{Z6spec_novs}}
\end{table}

\subsection{$\bZ_6'$, $v={1\over 6}(1,-3,2)$}
\label{Z6'}

It is possible to put all the $D5_3$-branes at the origin of the first 
two tori. The tadpole cancellation conditions for this configuration are:
\bea
k=0: &\ &\Tr(\gamma_{0,9}) = \Tr(\gamma_{0,5}) = 32 \nonumber\\
k=1: &  &\Tr(\gamma_{1,9})+2\,\Tr(\gamma_{1,5}) = 0,\qquad
         \Tr(\gamma_{1,9}) = 0 \nonumber\\
k=2: &  &\Tr(\gamma_{2,9}) = 8,\qquad
         \Tr(\gamma_{2,5}) = 8 \\
k=3: &  &\Tr(\gamma_{3,9})-4\,\Tr(\gamma_{3,5}) = 0,\qquad
         \Tr(\gamma_{3,9}) = 0 \nonumber
\eea
There is a unique solution to these equations, leading to the spectrum shown
in table \ref{Z6'spec}. This model is selfdual under T-duality in the first
two tori. As expected, the conditions for \nonab anomaly freedom are 
equivalent to the tadpole conditions for the sectors $k=1,2$.

\begin{table}[h]
\renewcommand{\arraystretch}{1.25}
$$\begin{array}{|c|c|}
\hline
\multicolumn{2}{|c|}{\IZ_6',\ v=\frac16(1,-3,2),\ \mu=1}\\ \hline\hline
\multicolumn{2}{|c|}{\hbox{open string spectrum}}\\ \hline
\rm sector &\hbox{gauge group / matter fields}\\ \hline
99    &SO(8)_1\times U(4)_2\times U(4)_3\times SO(8)_4\\
      &(\Yfun_1,\Yfunb_2),\ (\Yfun_1,\Yfunb_3),\ (\Yfun_1,\Yfun_4),\ 
       (\Yfun_2,\Yfun_3),\ (\Yfun_2,\Yfunb_3),\ (\Yfunb_2,\Yfunb_3)\\
      &(\Yfun_2,\Yfun_4),\ (\Yfun_3,\Yfun_4),\ \Yasym_3,\ \Yasymb_2\\ \hline
55    &\hbox{identical to 99}\\ \hline
95    &(\Yfun_1,\Yfunb_2),\ (\Yfunb_2,\Yfun_1),\ (\Yfun_2,\Yfunb_3),\ 
       (\Yfunb_3,\Yfun_2),\ (\Yfun_3,\Yfun_4),\ (\Yfun_4,\Yfun_3)\\
\hline\hline
\multicolumn{2}{|c|}{\hbox{closed string spectrum}}\\ \hline
\rm sector  &\hbox{$\cN=1$ multiplets}\\ \hline
\rm untw.   & \hbox{gravity, 1 lin., 4 chir.} \\ \hline
k=N/2       & \hbox{8 lin., 4 vec.} \\ \hline
0<k<N/2     & \hbox{18 lin., 9 chir., 3 vec.} \\ \hline
\end{array}$$
\caption{Spectrum of the $\bZ_6'$ orientifold with vector structure.
         \label{Z6'spec}}
\end{table}

For completeness and for comparison, we also give the spectrum of the
$\IZ_6'$ orientifold without vector structure and all $D5_3$-branes at
the origin, table \ref{Z6'spec_novs}. 
This orientifold has been analysed by the authors of \cite{z,afiv,abiu}.
The solution to the tadpole equations is unique and selfdual under T-duality 
in the first two tori. 

\begin{table}[htp]
\renewcommand{\arraystretch}{1.25}
$$\begin{array}{|c|c|}
\hline
\multicolumn{2}{|c|}{\IZ_6',\ v=\frac16(1,-3,2),\ \mu=-1}\\ \hline\hline
\multicolumn{2}{|c|}{\hbox{open string spectrum}}\\ \hline
\rm sector &\hbox{gauge group / matter fields}\\ \hline
99    &U(4)_1\times U(8)_2\times U(4)_3\\
      &(\Yfun_1,\Yfunb_2),\ (\Yfunb_1,\Yfunb_2),\ 
       (\Yfun_1,\Yfun_3),\ (\Yfun_1,\Yfunb_3),\ (\Yfunb_1,\Yfunb_3),\\
      &(\Yfun_2,\Yfun_3),\ (\Yfun_2,\Yfunb_3),\ 
       \Yasymb_1,\ \Yasym_2,\ \Yasymb_2,\ \Yasym_3 \\ \hline
55    &\hbox{identical to 99}\\ \hline
95    &(\Yfun_1,\Yfun_1),\ (\Yfunb_1,\Yfun_2),\ (\Yfun_2,\Yfunb_1),\ 
       (\Yfunb_2,\Yfun_3),\ (\Yfun_3,\Yfunb_2),\ (\Yfunb_3,\Yfunb_3)\\ 
\hline\hline
\multicolumn{2}{|c|}{\hbox{closed string spectrum}}\\ \hline
\rm sector  &\hbox{$\cN=1$ multiplets}\\ \hline
\rm untw.   & \hbox{gravity, 1 lin., 5 chir.} \\ \hline
k=N/2       & \hbox{12 chir.} \\ \hline
0<k<N/2     & \hbox{18 lin., 9 chir., 3 vec.} \\ \hline
\end{array}$$
\caption{Spectrum of the $\bZ_6'$ orientifold without vector structure.
         \label{Z6'spec_novs}}
\end{table}

\subsection{$\bZ_8$, $v={1\over8}(1,3,-4)$}
\label{Z8}

The construction of this orientifold model is more complicated. We will see 
that it is not possible to satisfy the tadpole conditions if all $D5_3$-branes
are put at the origin. Let us first analyse the fixed points of the twisted
sectors. The lattice corresponding to the first two complex planes cannot 
be decomposed in two two-dimensional sublattices.\footnote{The 
compactification lattice of the $\bZ_8$ orientifold is the root lattice of 
the Lie algebra $B_4\times D_2$.} Nevertheless, the Lefschetz
formula for the number of fixed points is still valid, when the above shift
vector is used. One finds two $\IZ_8$ fixed points in the first two complex
planes, one of them is the origin. In the $k=2$ sector there are four
$\IZ_4$ fixed points (more precisely, fixed tori extended in the third complex
plane). Two of them are also fixed under $\IZ_8$, the other two
are permuted under the action of $\IZ_8$. The $k=3$ sector only has the two
$\IZ_8$ fixed points and in the $k=4$ sector we find 16 $\IZ_2$ fixed points.

We want to find the simplest brane configuration that satisfies all the tadpole
conditions. The 32 $D5_3$-branes are divided into several sets. Each of these
sets of branes is located at a different point in the first two complex planes.
We denote the two $\IZ_8$ fixed points by the index $i=0,1$, the two $\IZ_4$
fixed points which are not fixed under $\IZ_8$ by the index $J=2,3$ and the
12 $\IZ_2$ fixed points which are not fixed under $\IZ_4$ by the index 
$I=4,\ldots,15$. Then the tadpole cancellation conditions are:
\bea
k=0: &\ &\Tr(\gamma_{0,9}) = 
         \sum_i\Tr(\gamma_{0,5,i})+\sum_J\Tr(\gamma_{0,5,J})
         +\sum_I\Tr(\gamma_{0,5,I}) = 32 \nonumber\\
k=1: &  &\Tr(\gamma_{1,9})-\sqrt2\,\Tr(\gamma_{1,5,i}) = 0 \nonumber\\
k=2: &  &\Tr(\gamma_{2,9})-2\,\Tr(\gamma_{2,5,i})  = 0,\qquad
         \Tr(\gamma_{2,9})-2\,\Tr(\gamma_{2,5,J})  = 0 \\
k=3: &  &\Tr(\gamma_{3,9})+\sqrt2\,\Tr(\gamma_{3,5,i}) = 0 \nonumber\\
k=4: &  &\Tr(\gamma_{4,9})+4\,\Tr(\gamma_{4,5,i})  = 32,\qquad
         \Tr(\gamma_{4,9})+4\,\Tr(\gamma_{4,5,J})  = 32,\nonumber\\
     &  &\Tr(\gamma_{4,9})+4\,\Tr(\gamma_{4,5,I})  = 0 \nonumber
\eea
Note that there is a \nonvan contribution of the Klein bottle to the
four $\IZ_4$ fixed points in the $k=4$ sector. The contribution of 
$\cK_0(2)$ and $\cK_1(2)$ is not present at the other 12 $\IZ_2$ fixed
points because the twists $k'=k/2=2$ and $k'=k/2+N/2=6$ only leave the
$\IZ_4$ fixed points invariant. This in contrast to the models without 
vector structure, where the Klein bottle contribution to the $k=N/2$ 
sector always vanishes at all fixed points. 

To cancel the tadpoles from the $k=4$ sector, some branes need to be
located at the two $\IZ_8$ fixed points $i=0,1$ and at the pair of 
$\IZ_4$ fixed points $J=2,3$ (the branes at the two points $J=2,3$ must
be identical to be $\IZ_8$-invariant). The tadpole equations have many
solutions. The most symmetric one corresponds to a configuration with 
8 $D5_3$-branes at each of the four $\IZ_4$ fixed points. There are
still 25 different solutions for the $\gamma$ matrices. The gauge groups 
corresponding to these solutions depend on the parameters $m=0,\ldots,4$,
$n=0,\ldots,8$, satisfying the conditions $m+n\le8$ and $m\le n$.
In table \ref{Z8spec}, we give the complete spectrum. 
We denoted the different sets of $D5_3$-branes by $5_0$, $5_1$, $5_J$, 
referring to the fixed point at the origin, the second $\IZ_8$ fixed point 
and the pair of $\IZ_4$ fixed points which are permuted under the action 
of $\IZ_8$. Note that the spectrum of the $5_J5_J$ sector is that of a 
$\IZ_4$ orientifold with shift vector $v=\frac14(1,3,-4)$.

\begin{table}[htp]
\renewcommand{\arraystretch}{1.25}
$$\begin{array}{|c|c|}
\hline
\multicolumn{2}{|c|}{\IZ_8,\ v=\frac18(1,3,-4),\ \mu=1}\\ \hline\hline
\multicolumn{2}{|c|}{\hbox{open string spectrum}}\\ \hline
\rm sector &\hbox{gauge group / matter fields}\\ \hline
99    &SO(8-2m)_1\times U(8-n)_2\times U(2m)_3\times U(n)_4\times 
       SO(8-2m)_5\\
      &(\Yfun_1\Yfunb_2),\ (\Yfun_1,\Yfunb_4),\ (\Yfun_1,\Yfun_5),\  
       (\Yfun_2,\Yfunb_3),\ (\Yfunb_2,\Yfunb_3),\ (\Yfun_2,\Yfun_4)\\
      & (\Yfunb_2,\Yfunb_4),\ (\Yfun_2,\Yfun_5),\ (\Yfun_3,\Yfun_4),\ 
        (\Yfun_3,\Yfunb_4),\ (\Yfun_4,\Yfun_5),\ \Yasym_3,\ \Yasymb_3\\ \hline
5_05_0&SO(8-m-n)_1\times U(m)_3\times SO(n-m)_5 \\ 
      &(\Yfun_1,\Yfun_5),\ \Yasym_3,\ \Yasymb_3\\ \hline  
5_15_1&\hbox{identical to $5_05_0$}\\ \hline
5_J5_J&SO(8-2m)_1\times SO(2m)_3\\
      &\Yasym_1,\ \Yasym_3\\ \hline
95_0    &(\Yfun_1,\Yfun_3),\ (\Yfun_3,\Yfun_1),\ 
         (\Yfunb_3,\Yfun_5),\ (\Yfun_5,\Yfunb_3)\\ \hline
95_1    &\hbox{identical to $95_0$}\\ \hline
95_J    &(\Yfun_1,\Yfun_3),\ (\Yfun_3,\Yfun_1),\ (\Yfunb_3,\Yfun_1),\ 
         (\Yfun_5,\Yfun_3)\\
\hline\hline
\multicolumn{2}{|c|}{\hbox{closed string spectrum}}\\ \hline
\rm sector  &\hbox{$\cN=1$ multiplets}\\ \hline
\rm untw.   & \hbox{gravity, 1 lin., 4 chir.} \\ \hline
k=N/2       & \hbox{6 lin., 4 vec.} \\ \hline
0<k<N/2     & \hbox{19 lin., 4 chir., 1 vec.} \\ \hline
\end{array}$$
\caption{Spectrum of the $\bZ_8$ orientifold with vector structure. The 
         parameters $m=0,\ldots,4$, $n=0,\ldots,8$ satisfy $m+n\le8$
         and $m\le n$. Some of the possible gauge group factors are
         missing (\eg the second and fourth factor in $5_i5_i$ sectors). This
         is due to the tadpole cancellation conditions which force the rank of
         these group factors to vanish.
         \label{Z8spec}}
\end{table}

The tadpole conditions of the sectors $k=1,3$ are equivalent the conditions
that can be derived from the requirement of anomaly freedom. Note that the
tadpole equations of the $k=4$ sector forced us to put $D5_3$-branes at 
different fixed points. However, these conditions are not necessary for
anomaly cancellation. Indeed, it would be possible to construct an 
anomaly free $\IZ_8$ orientifold with all $D5_3$-branes sitting at the 
origin. In the non-compact limit, such a model is consistent.

\subsection{$\bZ_8'$, $v={1\over8}(1,-3,2)$}
\label{Z8'}

We did not find a consistent solution for the $\IZ_8'$ orientifold. Below
we show the tadpole cancellation conditions and the general spectrum of
$D$-branes at a $\IZ_8'$ singularity. The fixed points in the first two complex
planes are identical to the fixed points of the $\IZ_8$ model.\footnote{The 
compactification lattice of the $\bZ_8'$ orientifold is the root lattice of 
the Lie algebra $B_4\times B_2$, \ie the four-dimensional sublattice 
corresponding to the first two complex planes is not factorisable. However, 
the Lefschetz formula for the number of fixed points is still valid.} The 
only difference is that now the third complex plane is only fixed under the
$k=4$ twist. There are two $\IZ_8$ fixed points in the first two complex
planes, one of them is the origin. In the $k=2$ sector there 
are four $\IZ_4$ fixed points. Two of them are also fixed under $\IZ_8$, 
the other two are permuted under the action of $\IZ_8$. The $k=3$ sector 
only has the two $\IZ_8$ fixed points and in the $k=4$ sector we find 16 
$\IZ_2$ fixed points (more precisely, fixed tori extended in the third
complex plane). As for the previous model, we denote the two $\IZ_8$ fixed 
points by the index $i=0,1$, the two $\IZ_4$ fixed points which are not 
fixed under $\IZ_8$ by the index $J=2,3$ and the 12 $\IZ_2$ fixed points 
which are not fixed under $\IZ_4$ by the index $I=4,\ldots,15$. Then the 
tadpole cancellation conditions are:
\bea 
k=0: &\ &\Tr(\gamma_{0,9}) = 
         \sum_i\Tr(\gamma_{0,5,i})+\sum_J\Tr(\gamma_{0,5,J})
         +\sum_I\Tr(\gamma_{0,5,I}) = 32 \nonumber\\
k=1: &  &\Tr(\gamma_{1,9})+\sqrt2\,\Tr(\gamma_{1,5,i}) = 0 \nonumber\\
k=2: &  &\Tr(\gamma_{2,9})+2\,\Tr(\gamma_{2,5,i})  = 16,\qquad
         \Tr(\gamma_{2,9})+2\,\Tr(\gamma_{2,5,J})  = 0 \nonumber\\
k=3: &  &\Tr(\gamma_{3,9})-\sqrt2\,\Tr(\gamma_{3,5,i}) = 0 \\
k=4: &  &\Tr(\gamma_{4,9})-4\,\Tr(\gamma_{4,5,i})  = 0,\qquad
         \Tr(\gamma_{4,9})-4\,\Tr(\gamma_{4,5,J})  = 0, \nonumber\\
       &&\Tr(\gamma_{4,9})-4\,\Tr(\gamma_{4,5,I})  = 0 \nonumber
\eea
From the first condition of the $k=2$ twisted sector, we can see that some 
$D5_3$-branes have to sit at each of the two $\IZ_8$ fixed points to cancel 
the tadpoles. This is the only orientifold where the necessity to put 
$D5_3$-branes at different fixed points arises due to the tadpole conditions
of a sector without fixed planes. Equivalently it can be verified that the
general spectrum of a configuration with all $D5_3$-branes at the origin
has no solution for the $\gamma$ matrices such that the spectrum is free
of \nonab gauge anomalies. 

The general spectrum for a configuration with two sets of $D5_3$-branes located
at the two $\IZ_8$ fixed points $i=0,1$, is shown in table \ref{Z8'spec}.
The numbers $n_j$, $m_j$, $l_j$ should be fixed by the tadpole conditions.
We must have overlooked some subtlety because it turns out that the conditions
for anomaly freedom of $U(n_2)$, $U(n_4)$, $U(m_2)$, $U(m_4)$, $U(l_2)$ and
$U(l_4)$ lead to $\Tr(\gamma_{2,9})=0$, $\Tr(\gamma_{2,5,i})=4$. This is
incompatible with the tadpole equations of the $k=2$ sector.

\begin{table}[htp]
\renewcommand{\arraystretch}{1.25}
$$\begin{array}{|c|c|}
\hline
\multicolumn{2}{|c|}{\IZ_8',\ v=\frac18(1,-3,2),\ \mu=1}\\ \hline\hline
\multicolumn{2}{|c|}{\hbox{open string spectrum}}\\ \hline
\rm sector &\hbox{gauge group / matter fields}\\ \hline
99    &SO(n_1)\times U(n_2)\times U(n_3)\times U(n_4)\times SO(n_5)\\
      &(\Yfun_1,\Yfunb_2),\ (\Yfun_1,\Yfunb_3),\ (\Yfun_1,\Yfun_4),\ 
       (\Yfun_2,\Yfun_3),\ (\Yfun_2,\Yfunb_3),\ (\Yfun_2,\Yfunb_4),\ 
       (\Yfunb_2,\Yfun_5)\\
      &(\Yfun_3,\Yfunb_4),\ (\Yfunb_3,\Yfunb_4),\ (\Yfun_3,\Yfun_5),\ 
       (\Yfun_4,\Yfun_5),\ \Yasymb_2,\ \Yasym_4  \\ \hline
5_05_0&SO(m_1)\times U(m_2)\times U(m_3)\times U(m_4)\times SO(m_5)\\
      &\hbox{matter identical to $99$}\\ \hline
5_15_1&SO(l_1)\times U(l_2)\times U(l_3)\times U(l_4)\times SO(l_5)\\
      &\hbox{matter identical to $99$}\\ \hline
95_0    &(\Yfun_1,\Yfunb_2),\ (\Yfunb_2,\Yfun_1),\ (\Yfun_2,\Yfunb_3),\ 
         (\Yfunb_3,\Yfun_2),\ (\Yfun_3,\Yfunb_4),\ (\Yfunb_4,\Yfun_3)\\
        &(\Yfun_4,\Yfun_5),\ (\Yfun_5,\Yfun_4)\\ \hline
95_1    &\hbox{identical to $95_0$}\\
\hline\hline
\multicolumn{2}{|c|}{\hbox{closed string spectrum}}\\ \hline
\rm sector  &\hbox{$\cN=1$ multiplets}\\ \hline
\rm untw.   & \hbox{gravity, 1 lin., 3 chir.} \\ \hline
k=N/2       & \hbox{6 lin., 3 vec.} \\ \hline
0<k<N/2     & \hbox{18 lin.} \\ \hline
\end{array}$$
\caption{Spectrum of the $\bZ_8'$ orientifold with vector structure. The
         parameters $n_j$, $m_j$, $l_j$ should be fixed by solving the
         tadpole cancellation conditions.
         \label{Z8'spec}}
\end{table}

\subsection{$\bZ_{12}$, $v={1\over{12}}(1,-5,4)$}
\label{Z12}

It is not possible to locate all the $D5_3$-branes at the origin.
In this case, the tadpole cancellation conditions of the order-two sector 
$k=6$ are responsible for this (i.e.\  it is not related to \nonab anomaly 
cancellation). This is analogous to the case of the $\IZ_8$ orientifold.
In the the sectors $k=1,2,5$ there is only fixed point in 
the first two complex planes\footnote{The compactification lattice of the 
$\bZ_{12}$ orientifold is the root lattice of the Lie algebra $F_4\times A_2$,
\ie the four-dimensional sublattice corresponding to the first two complex 
planes is not factorisable. However, the Lefschetz formula for the number of 
fixed points is still valid.}: the origin. In the $k=3$ sector there are four 
$\IZ_4$ fixed points (more precisely, four fixed tori extended in the third 
complex plane). In the $k=4$ sector there are 9 $\IZ_3$ fixed points in the 
first two complex planes and in the $k=6$ sector 16 $\IZ_2$ fixed points.
We denote the origin by $0$, the three $\IZ_4$ fixed points which are not 
fixed under $\IZ_{12}$ by the index $i=1,2,3$, the 8 $\IZ_3$ fixed points
which are not fixed under $\IZ_{12}$ by the index $J=1,\ldots,8$ and the 12
$\IZ_2$ fixed points which are not fixed under $\IZ_4$ by the index 
$I=4,\ldots,15$. Then the tadpole cancellation conditions are:
\bea 
k=0: &\ &\Tr(\gamma_{0,9}) = \Tr(\gamma_{0,5,0})
         +\sum_i\Tr(\gamma_{0,5,i})+\sum_J\Tr(\gamma_{0,5,J})
         +\sum_I\Tr(\gamma_{0,5,I}) = 32 \nonumber\\
k=1: &  &\Tr(\gamma_{1,9}) +  \Tr(\gamma_{1,5,0}) = 0 \nonumber\\
k=2: &  &\Tr(\gamma_{2,9}) + \Tr(\gamma_{2,5,0})  = 8 \\
k=3: &  &\Tr(\gamma_{3,9}) - 2\,\Tr(\gamma_{3,5,0}) = 0,\qquad
         \Tr(\gamma_{3,9}) - 2\,\Tr(\gamma_{3,5,i}) = 0 \nonumber\\
k=4: &  &\Tr(\gamma_{4,9}) - 3\,\Tr(\gamma_{4,5,0})  = 8,\qquad
         \Tr(\gamma_{4,9}) - 3\,\Tr(\gamma_{4,5,J})  = -4 \nonumber\\
k=5: &  &\Tr(\gamma_{5,9}) + \Tr(\gamma_{5,5,0}) = 0 \nonumber\\
k=6: &  &\Tr(\gamma_{6,9}) + 4\,\Tr\gamma_{6,5,0}  = 32,\qquad
         \Tr(\gamma_{6,9}) + 4\,Tr\gamma_{6,5,i}  = 32, \nonumber\\
       &&\Tr(\gamma_{6,9}) + 4\,Tr\gamma_{6,5,I}  = 0 \nonumber
\eea

In order to cancel the $k=6$ twisted tadpoles there must be some $D5_3$-branes
at each $\IZ_4$ fixed point. One of these is the origin and the other three 
are permuted by the  $\IZ_{12}$ action. Thus the branes at the three points 
$i=1,2,3$ must be identical to be $\IZ_{12}$-invariant. The tadpole equations 
have many solutions. The most symmetric one corresponds to a configuration 
with 8 $D5_3$-branes at each of the four $\IZ_4$ fixed points. There are
still 5 different solutions for the $\gamma$ matrices. 
In table \ref{Z12spec}, we give the complete spectrum. 
We denoted the different sets of $D5_3$-branes by by $5_0$, $5_i$, 
referring to the fixed point at the origin and the three $\IZ_4$ fixed points 
which are permuted under the action of $\IZ_{12}$.
Note that the spectrum of the $5_i5_i$ sector is that of a $\IZ_4$ 
orientifold with shift vector $v=\frac14(1,5,-4)$.

\begin{table}[htp]
\renewcommand{\arraystretch}{1.25}
$$\begin{array}{|c|c|}
\hline
\multicolumn{2}{|c|}{\IZ_{12},\ v=\frac1{12}(1,-5,4),\ \mu=1}\\ \hline\hline
\multicolumn{2}{|c|}{\hbox{open string spectrum}}\\ \hline
\rm sector &\hbox{gauge group / matter fields}\\ \hline
99    &SO(2n)_1\times U(4)_2\times U(4-n)_3\times U(n)_5\times
         U(4)_6\times SO(8-2n)_7\\
      &(\Yfun_1,\Yfunb_2),\ (\Yfun_1,\Yfunb_5),\ (\Yfun_1,\Yfun_6),\ 
       (\Yfun_2,\Yfunb_3),\ (\Yfun_2,\Yfun_5),\ (\Yfun_2,\Yfunb_6)\\ 
      &(\Yfunb_2,\Yfun_7),\ (\Yfunb_3,\Yfunb_6),\ (\Yfun_3,\Yfun_7),\ 
       (\Yfun_5,\Yfunb_6),\ (\Yfun_6,\Yfun_7),\ \Yasymb_3 ,\ \Yasym_5\\ \hline
5_05_0&U(4-n)_3\times U(n)_5 \\
      &\Yasymb_3,\ \Yasym_5\\ \hline 
5_i5_i&SO(2n)_1\times SO(8-2n)_3\\
      &\Yasym_1,\ \Yasym_3\\ \hline
95_0    &(\Yfun_1,\Yfunb_3),\ (\Yfun_3,\Yfunb_5),\ (\Yfunb_5,\Yfun_3),\ 
         (\Yfun_7,\Yfun_5)\\ \hline
95_i    &(\Yfun_7,\Yfun_1),\ (\Yfun_1,\Yfun_3)\\
\hline\hline
\multicolumn{2}{|c|}{\hbox{closed string spectrum}}\\ \hline
\rm sector  &\hbox{$\cN=1$ multiplets}\\ \hline
\rm untw.   & \hbox{gravity, 1 lin., 3 chir.} \\ \hline
k=N/2       & \hbox{4 lin., 3 vec.} \\ \hline
0<k<N/2     & \hbox{20 lin., 3 chir., 1 vec.} \\ \hline
\end{array}$$
\caption{Spectrum of the $\bZ_{12}$ orientifold with vector structure. There
         are 5 solutions parametrised by $n=0,\ldots,4$. The difference 
         between the 99 and the $5_05_0$ sector is due to the tadpole
         cancellation conditions which force many of the possible group
         factors to vanish.
         \label{Z12spec}}
\end{table}

We verified that the requirement of anomaly cancellation is equivalent to 
the twisted tadpole equations from the sectors $k=1,2,4,5$. The tadpole
conditions from the $k=6$ sector, which require to distribute the 
$D5_3$-branes over different fixed points, are not necessary for anomaly 
cancellation. One could, in principle, put all the $D5_3$-branes at the 
origin and produce a model free of anomalies. 

These models are not selfdual under T-duality because the $D5_3$-branes are 
at several fixed points. The dual model will require Wilson lines in the 
$D9$-brane sector. 

The orientifold without vector structure has been constructed in 
\cite{afiv,abiu}. There are 235 models without vector structure if 
all the $D5_3$-branes sit at the origin. For completeness and for 
comparison we display the spectrum in table \ref{Z12spec_novs}.
\begin{table}[htp]
\renewcommand{\arraystretch}{1.25}
$$\begin{array}{|c|c|}
\hline
\multicolumn{2}{|c|}{\IZ_{12},\ v=\frac1{12}(1,-5,4),\ \mu=-1}\\ \hline\hline
\multicolumn{2}{|c|}{\hbox{open string spectrum}}\\ \hline
\rm sector &\hbox{gauge group / matter fields}\\ \hline
99    &U(l)_1\times U(m)_2\times U(n)_3\times U(4+m-l)_4\times 
              U(4-m)_5\times U(8-m-n)_6\\
      &(\Yfun_1,\Yfunb_2),\ (\Yfunb_1,\Yfunb_4),\ (\Yfun_1,\Yfun_5),\ 
       (\Yfun_1,\Yfunb_5),\ (\Yfun_2,\Yfunb_3),\ (\Yfunb_2,\Yfunb_3)\\ 
      &(\Yfun_2,\Yfun_4),\ (\Yfun_2,\Yfunb_6),\ (\Yfunb_2,\Yfunb_6),\  
       (\Yfun_3,\Yfunb_4),\ (\Yfunb_3,\Yfunb_5),\ (\Yfun_3,\Yfun_6)\\ 
      &(\Yfun_4,\Yfun_5),\ (\Yfun_4,\Yfunb_5),\ (\Yfun_5,\Yfunb_6),\ 
       (\Yfun_6,\Yfunb_1),\ 
       \Yasymb_1 ,\ \Yasym_3\ \Yasymb_4 ,\ \Yasym_6\\ \hline
55    &\hbox{identical to 99}\\ \hline 
95    &(\Yfun_1,\Yfun_2),\ (\Yfunb_1,\Yfun_3),\ (\Yfun_2,\Yfun_1),\ 
       (\Yfunb_2,\Yfun_4),\ (\Yfun_3,\Yfunb_1),\ (\Yfunb_3,\Yfun_5)\\
      &(\Yfun_4,\Yfunb_2),\ (\Yfunb_4,\Yfun_6),\ (\Yfun_5,\Yfunb_3),\ 
       (\Yfunb_5,\Yfunb_6),\ (\Yfun_6,\Yfunb_4),\ (\Yfunb_6,\Yfunb_5)\\
\hline\hline
\multicolumn{2}{|c|}{\hbox{closed string spectrum}}\\ \hline
\rm sector  &\hbox{$\cN=1$ multiplets}\\ \hline
\rm untw.   & \hbox{gravity, 1 lin., 3 chir.} \\ \hline
k=N/2       & \hbox{7 chir.} \\ \hline
0<k<N/2     & \hbox{20 lin., 3 chir., 1 vec.} \\ \hline
\end{array}$$
\caption{Spectrum of the $\bZ_{12}$ orientifold without vector structure. 
         There are 235 solutions parametrised by $l,n=0,\ldots,8$, 
         $m=0,\ldots,4$, satisfying $l-m\le4$ and $n+m\le8$. 
         \label{Z12spec_novs}}
\end{table}

\subsection{$\bZ'_{12}$, $v={1\over{12}}(1,5,-6)$}
\label{Z12'}

This orientifold can be consistently constructed with all $D5_3$-branes 
sitting at the origin of the first two complex planes.\footnote{The 
compactification lattice of the $\bZ_{12}'$ orientifold is the root 
lattice of the Lie algebra $F_4\times D_2$, \ie the four-dimensional 
sublattice corresponding to the first two complex planes is not factorisable. 
However, the Lefschetz formula for the number of fixed points is still valid.}
In the $k=1,2,5$ sectors, only the origin is fixed.
In the  sector $k=3$, there are four fixed tori, one of them is $\IZ_{12}$ 
invariant (the origin) and the other form a triplet of $\IZ_4$ fixed points 
permuted by the $\IZ_{12}$ element. The $k=4$ twisted sector has 9 fixed 
points in the first two complex planes. 
The tadpole cancellation conditions for all $D5_3$-branes at the origin are:
\bea
k=0: &\ &\Tr(\gamma_{0,9}) = \Tr(\gamma_{0,5}) = 32 \nonumber\\
k=1: &  &\Tr(\gamma_{1,9}) - \Tr(\gamma_{1,5}) = 0 \nonumber\\
k=2: &  &\Tr(\gamma_{2,9}) - \Tr(\gamma_{2,5})  = 0 \nonumber\\
k=3: &  &\Tr(\gamma_{3,9}) + 2\,\Tr(\gamma_{3,5}) = 0,\qquad
         \Tr(\gamma_{3,9}) = 0 \nonumber\\
k=4: &  &\Tr(\gamma_{4,9}) + 3\,\Tr(\gamma_{4,5})  = 32,\qquad
         \Tr(\gamma_{4,9})  = 8 \\
k=5: &  &\Tr(\gamma_{5,9}) - \Tr(\gamma_{5,5}) = 0 \nonumber\\
k=6: &  &\Tr(\gamma_{6,9}) - 4\,\Tr(\gamma_{6,5})  = 0,\qquad
         \Tr(\gamma_{6,9})  = 0 \nonumber
\eea
There are 125  solutions to the tadpole cancellation conditions, table
\ref{Z12'spec}. All of them are selfdual under T-duality.

\begin{table}[htp]
\renewcommand{\arraystretch}{1.25}
$$\begin{array}{|c|c|}
\hline
\multicolumn{2}{|c|}{\IZ_{12}',\ v=\frac1{12}(1,5,-6),\ \mu=1}\\ \hline\hline
\multicolumn{2}{|c|}{\hbox{open string spectrum}}\\ \hline
\rm sector &\hbox{gauge group / matter fields}\\ \hline
99    &SO(8+2m-2n)_1\times U(l)_2\times U(m)_3\times U(n)_4\\
      &\qquad\times U(n-m)_5\times U(8-n-l)_6\times SO(8-2m)_7\\
      &(\Yfun_1,\Yfunb_2),\ (\Yfun_1,\Yfunb_6),\ (\Yfun_1,\Yfun_7),\ 
       (\Yfun_2,\Yfunb_3),\ (\Yfunb_2,\Yfunb_5),\ (\Yfun_2,\Yfun_6)\\
      &(\Yfunb_2,\Yfunb_6),\ (\Yfun_2,\Yfun_7),\ (\Yfun_3,\Yfunb_4),\ 
       (\Yfunb_3,\Yfunb_4),\ (\Yfun_3,\Yfun_5),\ (\Yfunb_3,\Yfunb_5)\\
      &(\Yfun_3,\Yfun_6),\ (\Yfun_4,\Yfun_5),\ (\Yfun_4,\Yfunb_5),\ 
       (\Yfun_5,\Yfunb_6),\ (\Yfun_6,\Yfun_7),\ \Yasym_4,\ \Yasymb_4\\ \hline
55    &\hbox{identical to 99}\\ \hline 
95    &(\Yfun_1,\Yfun_4),\ (\Yfun_2,\Yfun_3),\ (\Yfunb_2,\Yfun_5),\ 
       (\Yfun_3,\Yfun_2),\ (\Yfunb_3,\Yfun_6),\ (\Yfun_4,\Yfun_1)\\ 
      &(\Yfunb_4,\Yfun_7),\ (\Yfun_5,\Yfunb_2),\ (\Yfunb_5,\Yfunb_6),\
       (\Yfun_6,\Yfunb_3),\ (\Yfunb_6,\Yfunb_5),\ (\Yfun_7,\Yfun_4)\\ 
\hline\hline
\multicolumn{2}{|c|}{\hbox{closed string spectrum}}\\ \hline
\rm sector  &\hbox{$\cN=1$ multiplets}\\ \hline
\rm untw.   & \hbox{gravity, 1 lin., 4 chir.} \\ \hline
k=N/2       & \hbox{4 lin., 2 vec.} \\ \hline
0<k<N/2     & \hbox{20 lin., 6 chir., 2 vec.} \\ \hline
\end{array}$$
\caption{Spectrum of the $\bZ_{12}'$ orientifold with vector structure.
         There are 125 solutions parametrised by $l,n=0,\ldots,8$,
         $m=0,\ldots,4$, satisfying $0\le n-m\le 4$ and $l+n\le8$.
         \label{Z12'spec}}
\end{table}

We verified that the requirement of anomaly cancellation is equivalent to 
the twisted tadpole conditions of the sectors $k=1,2,5$.

\section{Conclusions}
\label{conclusion}

We have constructed \D4, \N1 orientifolds with vector structure corresponding
to orbifold groups $\IZ_N$, $N$ even. We found a consistent 
solution to the tadpole equations for each orbifold group (except $\IZ_8'$) 
that leads to a \D4, \N1 orbifold of the heterotic string. Due to the fact 
that the $\gamma$-matrices represent the action of $\IZ_N$ on the Chan-Paton 
indices only projectively, there may appear various signs in the Klein bottle,
M\"obius strip and cylinder amplitude. We carefully included all these
signs in the tadpole computation and found new solutions. In general,
there are two \noneq $\IZ_N$ orientifolds with vector structure for
each even $N$: a \nonsusy model similar to the $\IZ_4$ constructed in
\cite{ads} and a supersymmetric model constructed in the present article.
For some orbifold groups, there are also consistent models without
vector structure. Supersymmetric $\IZ_N$ orientifolds without vector
structure have been constructed in \cite{ks2,afiv}. But their \nonsusy
analogues should also exist. 

The standard construction of $\IZ_N$ orientifolds (generalisations of the
GP model to \D4) leads to models without vector structure. They contain
one $O9^-$-plane, 16 $O5^-$-planes, 32 $D9$-branes and 32 $D5$-branes.
It turns out that models with vector structure are only possible if a
sign in the twisted Klein bottle contribution is flipped \cite{ads,au,aaads}.
This can be interpreted as replacing the $O5^-$-planes by $O5^+$-planes. 
One possibility to cancel the tadpoles of this new models is to introduce 
anti-$D5$-branes. This leads to \nonsusy orientifolds. Another possibility
consists in flipping a second sign: the coefficient of the 95 cylinder 
contribution to the tadpoles. In contrast to the models containing
antibranes, this is a global sign flip, it affects the RR and the NSNS
tadpoles. As a consequence, supersymmetry is preserved.
We have not yet fully understood the physical interpretation
of this second sign flip. But it is tempting to believe that it corresponds
to replacing $D5$-branes by $D5^-$-branes which have negative RR charge and
negative NSNS charge.

The models constructed in this article are very similar to the \nonsusy
$\IZ_N$ orientifolds of \cite{au,aaads}. The fermionic spectrum of the
latter coincides with the fermionic spectrum of our models. The bosonic
spectrum of the \nonsusy $\IZ_N$ orientifolds with vector structure can
easily be obtained. In the 55 sector, one has to replace the orthogonal
gauge group factors of our models by symplectic ones and the matter fields 
in the antisymmetric tensor representation by fields in the symmetric 
representation. For the remaining fields in the 55 sector, there is
no difference between fermions and bosons in the \nonsusy models. In the 
$95_3$ sector, the bosons are obtained by replacing the factor 
$e^{\pi iv_3}$ by $e^{-\pi i(v_1-v_2)}$ in (\ref{mixed_proj}) (see \eg
\cite{au}).

We thank A.\ Sagnotti for drawing our attention to a subtlety of the 
models constructed in this article that is not yet understood.
As explained in section \ref{interpretation}, the negative NSNS charge
has a strange consequence in the open string channel.
If one translates the closed string exchange between a $D9$-brane and a
$D5^-$-brane to the corresponding open string one-loop amplitude, the NS
contribution to the partition function appears with a negative sign and
the R contribution with a positive sign. This seems to violate the 
spin-statistics theorem. For models containing only 9-branes, it has been
shown in \cite{bs2} that the correct signs for the NS and R contributions
are necessary for higher-loop modular invariance. However, the situation
is less clear when different types of boundaries are involved.

Quiver diagrams are a very useful tool to compute orientifold spectra.
Based on the work of \cite{dm,bi}, we give general rules how to obtain 
the orientifold quiver for a given discrete group. It is straightforward 
to implement this algorithm in a computer algebra program. This enables 
us to compute orientifold spectra in a fast and efficient way.

\vskip3cm

\centerline{\bf Acknowledgements}
It is a pleasure to thank Luis~Ib\'a\~nez, Augusto Sagnotti and 
Angel Uranga for many helpful discussions. 
The work of M.K.\ is supported by a TMR network of the European Union, 
ref. FMRX-CT96-0090. The work of R.R.\ is supported by the MEC through 
an FPU Grant.

\newpage
\begin{center}\huge\bf Appendix \end{center}
\begin{appendix}

\section{Orientifold quivers}
\label{app_quiver}

In this appendix, we briefly review quiver diagrams as a tool to determine
the spectrum of type II orbifolds \cite{dm} and then explain how this is 
generalised to orientifold models following the ideas of \cite{bi}.%
\footnote{A detailed discussion of orbifold quivers can be found in \cite{hh}
and the references therein. For a mathematical introduction to quiver theory, 
see \eg \cite{i}.}

Consider a set of $N_p$ $Dp$-branes at a $\Gamma$ orbifold singularity,
where $\Gamma$ is some finite group. The action of $\Gamma$ on the 
Chan-Paton indices of the open strings ending on the $Dp$-branes is 
described by a (projective) representation $\gamma$ that associates a 
$(N_p\times N_p)$-matrix $\gamma_g$ to each element $g$ of $\Gamma$.
\bea  \label{def_gam}
\gamma:\ \Gamma &\longrightarrow &GL(N_p,\IC) \\
              g &\longmapsto     &\gamma_g \nonumber
\eea
In general, $\gamma$ can be decomposed in a direct sum of irreducible 
representations (irreps) $R_l$:
\be  \label{gam_rep}
   \gamma=\bigoplus_l n_lR_l,
\ee
where the notation $n_lR_l$ is short for $R_l\otimes\one_{n_l}$,
\ie $n_l$ is the number of copies of the irrep $R_l$ in $\gamma$. 
The action of $\Gamma$ on the internal $\IC^3$ is described by a
representation $R_{C^3}$:
\bea  \label{def_RC3}
\gamma:\ R_{C^3} &\longrightarrow &SU(3) \\
               g &\longmapsto     &R_{C^3}(g) \nonumber
\eea
We write $R_{C^3}=R^{(1)}_{C^3}\oplus R^{(2)}_{C^3}\oplus R^{(3)}_{C^3}$
(this decomposition is possible whenever $\Gamma$ is Abelian),
where $R^{(i)}_{C^3}$ corresponds to the action of $R_{C^3}$ on the $i\th$
coordinate of $\IC^3$. Then the projection of the Chan-Paton matrices
$\lambda^{(0)}$ (gauge fields) and $\lambda^{(i)}$ (matter fields) on
$\Gamma$-invariant states reads:
\be  \label{orbspec}
\lambda^{(0)}=\gamma_g\lambda^{(0)}\gamma_g^{-1},\qquad
\lambda^{(i)}=R^{(i)}_{C^3}(g)\,\gamma_g\lambda^{(i)}\gamma_g^{-1}.
\ee
The solution of these equations leads to the gauge group
\be  \label{orbgauge}
G\ =\ \prod_l U(n_l)
\ee
and matter fields
\be  \label{orbmat}
\sum_{i=1}^3\sum_{k,l} a^{(i)}_{kl}\,(\Yfun_k,\Yfunb_l),
\ee
where the coefficients $a^{(i)}_{kl}$ only take the values 0 or 1 and are
defined through
\be  \label{def_akl}
R^{(i)}_{C^3}\otimes R_k\ =\ \bigoplus_l a^{(i)}_{kl}\,R_l.
\ee
Solving this equation for $a^{(i)}_{kl}$, one finds
\be  \label{akl}
a^{(i)}_{kl}\ =\ {1\over|\Gamma|}\sum_{g\in\Gamma}R^{(i)}_{C^3}(g)
                  \tr(R_k(g))\tr(R_l(g^{-1})).
\ee

Quiver diagrams are a nice graphical representation of the orbifold spectrum.
To obtain the quiver corresponding to a $\Gamma$ orbifold:
\begin{itemize}
\item[-] determine all irreps $R_l$ of $\Gamma$ and associate a node $\bullet$
         to each irrep,
\item[-] calculate the coefficients $a^{(i)}_{kl}$ and draw an oriented link
         from the $k\th$ to the $l\th$ node \qlink{c}{c} if $a^{(i)}_{kl}=1$.
\end{itemize}   

\begin{figure}[htp]
\begin{center}
\epsffile{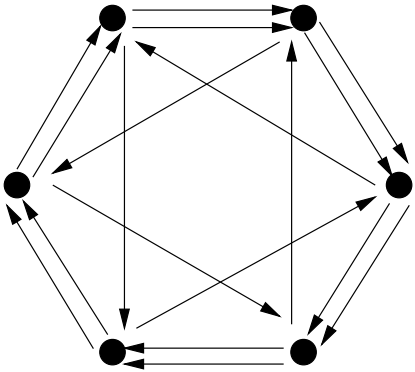}
\end{center}
\caption{\label{Z6quiver}Quiver diagram of the $\bZ_6$ orbifold with
         shift vector $v=\frac16(1,1,-2)$}
\end{figure}

For $\Gamma=\IZ_N$, there are $N$ irreps, shown in eq.\ (\ref{ZN_irreps}).
It is easy to see that the coefficients determining the matter representations
are $a^{(i)}_{kl}=\delta_{k,l+Nv_i\:{\rm mod\:}N}$, where $v=(v_1,v_2,v_3)$ 
is the shift vector defined in (\ref{shiftvec}). This leads to a quiver 
diagram similar to the one shown in figure \ref{Z6quiver}.

The orientifold model is obtained by projecting the corresponding orbifold
on $\Omega$-invariant states. As a consequence, the representation $\gamma$
of (\ref{def_gam}) must be real or pseudoreal. Again, $\gamma$ can be
decomposed in irreducible blocks, as in (\ref{gam_rep}). But now we have to 
distinguish between real, pseudoreal and complex irreps, eq.\ 
(\ref{gamma_rep}). Accordingly, we divide the set of indices $l$ that label
the irreps of $\Gamma$ in three subsets $\{l\}=\cR\cup\cP\cup\cC$.
Due to the $\Omega$-projection, the complex irreps contained in $\gamma$ 
only appear in pairs of conjugate representations, \ie if the $l\th$ irrep
is complex, then there exists an $\bar l$, such that $R_{\bar l}=\bar R_l$.
It is convenient to divide the complex irreps into two subsets 
$\cC=\cC_1\cup\cC_2$, $\cC_1\cap\cC_2=\emptyset$, such that for each
$l\in\cC_1$ the conjugate index $\bar l$ belongs to $\cC_2$ and vice versa.
In terms of quiver diagrams, this means that we have to identify nodes that 
correspond to conjugate representations. This leads to the following rules
for orientifold quivers: 

First draw a modified orbifold quiver diagram:
\begin{itemize}
\item[-] determine all irreps of $\Gamma$ and associate a node to each 
         irrep depending on whether it is real $\circ$, pseudoreal 
         \raisebox{0.5mm}{$\scriptstyle\odot$} or complex $\bullet$,
\item[-] compute the coefficients $a^{(i)}_{kl}$ using (\ref{akl}) and 
         for each $a^{(i)}_{kl}=1$ draw an oriented link from the $k\th$ 
         to the $l\th$ node, \eg \qlink{c}{c} if $k,l\in\cC$, 
         \qlink{r}{c} if $k\in\cR$, $l\in\cC$, etc. 
\end{itemize}

Then perform the $\Omega$-projection on this quiver diagram (in the following,
we only distinguish between $\circ$ and \raisebox{0.5mm}{$\scriptstyle\odot$}
if they are \noneq):
\begin{itemize}
\item[-] cancel the $l\th$ node if $l\in\cC_2$,
\item[-] cancel the link \qlink{c}{c} if 
         \begin{itemize}
         \item[$\cdot$] $k,l\in\cC_2$\ \ or
         \item[$\cdot$] $k\in\cC_1$, $l\in\cC_2$ and $k>\bar l$\ \ or
         \item[$\cdot$] $k\in\cC_2$, $l\in\cC_1$ and $\bar k>l$
         \end{itemize}
\item[-] replace the link \qlink{c}{c} by
         \begin{itemize}
         \item[$\cdot$] \qlinkx00cck{\bar l} 
                        if $k\in\cC_1$, $l\in\cC_2$, $k < \bar l$
         \item[$\cdot$] \qlinkx11cc{\bar k}l 
                        if $k\in\cC_2$, $l\in\cC_1$, $\bar k < l$
         \item[$\cdot$] \qtens{c}{k} if $k\in\cC_1$, $l\in\cC_2$, $k=\bar l$
         \item[$\cdot$] \qconjtens{l} if $k\in\cC_2$, $l\in\cC_1$, $\bar k=l$
         \end{itemize}
\item[-] cancel the link \qlink{r}{c} (or \qlinkx10rckl) if $l\in\cC_2$
\item[-] cancel the link \qlinkx01rrkl if $k>l$
\item[-] cancel all arrows (not the links) pointing towards a real node 
         $\circ$, \ie replace \qlinkx01crkl by \qlink{c}{r}, etc.
\end{itemize}

\begin{figure}[htp]
\begin{center}
\epsffile{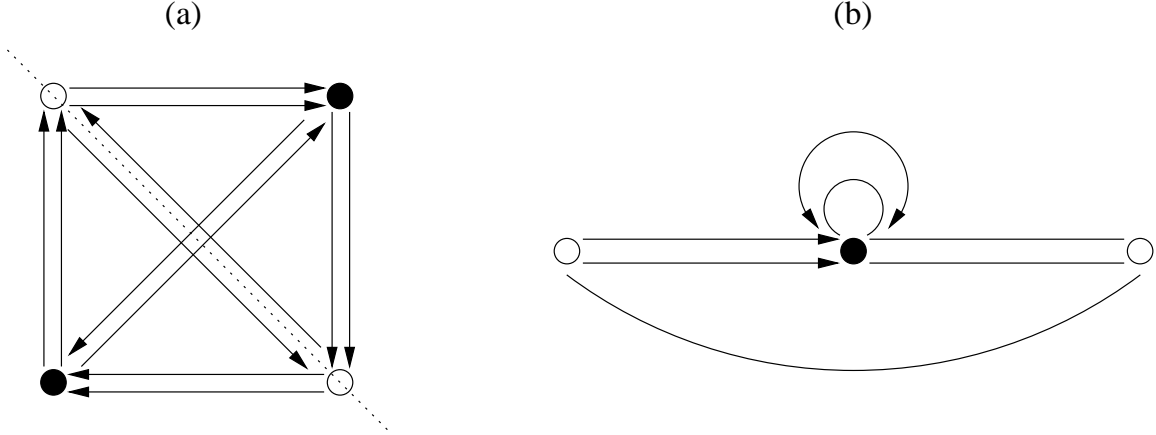}
\end{center}
\caption{\label{Z4quiver}Quiver diagram (a) of the $\bZ_4$ orbifold with
         shift vector $v=\frac14(1,1,-2)$ and (b) of the corresponding
         orientifold with vector structure. The dotted line indicates 
         the axis along which the $\Omega$-projection is performed.}
\end{figure}

In Figure \ref{Z4quiver} we illustrate these rules by showing how the
quiver diagram of the $\IZ_4$ orientifold with vector structure is obtained.

The spectrum can easily be read off from the orientifold quiver. Each node
corresponds to a gauge group factor:
$${\raisebox{-3mm}{$\bullet$}\atop\scriptstyle l}:\ U(n_l),\qquad
  {\raisebox{-3mm}{$\circ$}\atop\scriptstyle l}:\ 
           \left\{\ba{ll} SO(n_l) &{\rm if\ }c_p=1\\
                         USp(n_l) &{\rm if\ }c_p=-1
           \ea\right.,\qquad
  {\raisebox{-3mm}{$\scriptstyle\odot$}\atop\scriptstyle l}:\ 
           \left\{\ba{ll} USp(n_l) &{\rm if\ }c_p=1\\
                           SO(n_l) &{\rm if\ }c_p=-1
           \ea\right.,$$
where $c_p=+1\ (-1)$ if $\gamma_{\Omega,p}$ is (anti)symmetric.
Each link corresponds to a matter field transforming as a bifundamental
(if it connects two different nodes) or as a second rank tensor (if it starts
and ends on the same node). The representation is by definition fundamental
at the tail of the link and antifundamental at the head of the link.
$$\hbox{\qlink cc}:\ (\Yfun_k,\Yfunb_l),\qquad
  \hbox{\qlink cr}:\ (\Yfun_k,\Yfun_l),\qquad
  \hbox{\qlink rr}:\ (\Yfun_k,\Yfun_l),\qquad\ldots$$
$$\hbox{\qadj k}:\ adj_k,\qquad
  \hbox{\qtens ck}:\ \left\{\ba{ll} \Yasym_k &{\rm if\ }c_p=1\\
                                    \Ysym_k &{\rm if\ }c_p=-1
                     \ea\right.,\qquad
  \hbox{\qconjtens k}:\ \left\{\ba{ll} \Yasymb_k &{\rm if\ }c_p=1\\
                                    \Ysymb_k &{\rm if\ }c_p=-1
                        \ea\right.$$
It turns out that it is more difficult to find the tensor representations of 
orthogonal or symplectic groups. We need to define
\be  \label{R-index}
c^{(i)}_k\ =\ {1\over|\Gamma|}\sum_{g\in\Gamma}R^{(i)}_{C^3}(g)\tr(R_k(g^2)).
\ee
Then, one has
$$\hbox{\qtens rk}:\ \left\{\ba{ll} \Yasym_k &{\rm if\ }c_pc^{(i)}_k=1\\
                                    \Ysym_k &{\rm if\ }c_pc^{(i)}_k=-1
                     \ea\right.$$
Note that the index (\ref{R-index}) is a direct generalisation of the
well-known Frobenius-Schur index
\be  \label{FS-index}
c^{\rm FS}_k\ =\ {1\over|\Gamma|}\sum_{g\in\Gamma}\tr(R_k(g^2))\ = \ 
 \left\{\ba{ll} 1 &{\rm if\ }R_k\ {\rm real}\\
               -1 &{\rm if\ }R_k\ {\rm pseudoreal}\\
                0 &{\rm if\ }R_k\ {\rm complex}
 \ea\right.
\ee

Applying these rules to the quiver diagram of the $\IZ_4$ orientifold,
figure \ref{Z4quiver}, one immediately finds the spectrum of the 99 sector
and the 55 sector displayed in table \ref{Z4spec}.

The above rules to determine the orientifold spectrum are easily generalised
to also include the 95 sector. One has to calculate the coefficients
$a^{95}_{kl}$ defined by
\be  \label{akl_mixed}
a^{95}_{kl}\ =\ {1\over|\Gamma|}\sum_{g\in\Gamma}
                  \left(R^{(3)}_{C^3}(g)\right)^{1/2}
                  \tr(R_k(g))\tr(R_l(g^{-1})).
\ee
By the arguments given at the beginning of this appendix,
we find that the matter fields transforming in representations
\be  \label{mixedmat}
\sum_{k,l} a^{95}_{kl}\,(\Yfun_k,\Yfunb_l),
\ee
solve the projection equation (\ref{mixed_proj}). The first entry in the
bifundamental in (\ref{mixedmat}) corresponds to the transformation under
the 99 gauge group and the second entry to the transformation under the 55
gauge group. This is already the orientifold spectrum because $\Omega$ only
relates the 95 sector to the 59 sector but it imposes no new condition.
Of course, one has to replace $(\Yfun_k,\Yfunb_l)$ by 
$(\Yfunb_{\bar k},\Yfunb_l)$ if $k\in\cC_2$ and similarly if $l\in\cC_2$.

\section{Tadpoles of $\bZ_N$ orientifolds} 
\label{app_tad}
In this appendix, we use the strategy of \cite{kr1,pru} to derive the 
tadpole cancellation conditions for $\IZ_N$ orientifolds. The idea is 
that the low-energy limit ($t\to0$) of the one-loop amplitudes contributing 
to the tadpoles can be expressed as a sum over products of sines, cosines 
and traces of $\gamma$ matrices. This method can be directly applied to the 
\noncom orientifolds. Two modifications arise in the case of compact 
orientifold models: First, the zeros or singularities of some sine- or 
cosine-factor have to be replaced by the appropriate volume factor $V_i$ 
of the $i\th$ internal torus. Second, the compactification leads to new 
fixed points, where new contributions to the tadpoles may appear. We will 
not consider the volume dependence of the four \noncom coordinates because 
it can be trivially factorised out. 

In the following, we label the elements of $\IZ_N$ by the integer 
$k=0,\ldots,N-1$, and define $s_i=\sin(\pi kv_i)$, $c_i=\cos(\pi kv_i)$ 
and $\tilde s_i=\sin(2\pi kv_i)$.

The cylinder contribution to the tadpole cancellation conditions is:
\beq  \label{tadcyl}
\cC\ =\ \sum_{k=0}^{N-1}\cC_{(k)}\ =\ 
        \sum_{k=0}^{N-1} \frac{1}{8s_1s_2s_3} 
        \left[\Tr\gamma_{k,9} + 4\,\alpha\,s_1 s_2\Tr\gamma_{k,5}\right]^2,
\eeq
where $\alpha$ is a sign related to the weight of the $95_3$ sector. For
supersymmetric models, it has been shown in \cite{dp} that $\alpha=-1$ if 
one uses the alternative action of $\Omega^2$ on the oscillator ground state 
of the $95_3$ sector, $\Omega^2|_{95_3}=+1$, whereas $\alpha=+1$ for the 
standard GP action.\footnote{Non-supersymmetric models with GP action of
$\Omega^2$ on the $95_3$ sector have $\alpha=-1$ in the RR tadpoles.}
The formula (\ref{tadcyl}) is valid if there are no fixed tori. If the $i\th$ 
internal torus is fixed, \ie if $kv_i=0{\rm\ mod\ }\IZ$, one must take into 
account the volume dependence:
\begin{itemize}
\item if $kv_3=0{\rm\ mod\ }\IZ$:
\beq  \label{tadcyl3}
\cC_{(k)}\ =\ \frac{V_3}{8s_1s_2} 
           \left[\Tr\gamma_{k,9} + 4\,\alpha\,s_1 s_2\Tr\gamma_{k,5}\right]^2.
\eeq

\item if $kv_{i\neq3}=0{\rm\ mod\ }\IZ$:
\beq  \label{tadcyl12}
\cC_{(k)}\ =\ \mp\frac{V_i}{8s_3^2} [\Tr\gamma_{k,9}]^2
               \mp  \frac{2}{V_i} [\Tr\gamma_{k,5}]^2
               + \frac{\alpha}{s_3} \Tr\gamma_{k,9}\Tr\gamma_{k,5} ,
\eeq
where the upper (lower) sign refers to $kv_i=$even (odd).

\item if $k=0$:
\beq \label{tadcyluntw}
\cC_{(0)}\ =\ \frac{V_1V_2V_3}{8} [\Tr\gamma_{0,9}]^2 
            + \frac{2V_3}{V_1V_2} [\Tr\gamma_{0,5}]^2 
            + \alpha\,V_3 \Tr\gamma_{0,9}\Tr\gamma_{0,5} .
\eeq
\end{itemize}

The Klein bottle contribution can be split as:
\beq
\cK = \sum_{k=0}^{N-1}\left( \cK_0 ({k}) + \cK_1 ({k}) \right),
\eeq
where $\cK_0$ is the contribution of the untwisted sector and $\cK_1$ is 
the contribution of the order-two sector (see \cite{kr1,afiv}). 
One has:
\beq
\cK_0 ({k}) = 16 \prod_{i=1}^3 \frac{2 c_i^2}{\tilde{s_i}}, \qquad
\cK_1 ({k}) = 16\,\epsilon\,\frac{2 c_3^2}{\tilde s_3}
\eeq
The sign $\epsilon$ is related to the choice of the $\Omega$-projection.
One has $\epsilon = +1$ for the GP projection and $\epsilon = -1$ for the
DPBZ projection, eq.\ (\ref{OmJT}).
The Klein bottle contribution to the tadpole conditions can be reordered:
\beq
\cK =\sum_{k=0}^{N/2-1}\frac{1}{8\tilde{s}_1\tilde{s}_2\tilde{s}_3} 
                      [32\,(c_1c_2c_3 + \epsilon\, s_1 s_2 c_3)]^2.
\eeq

These formulae are valid if $2kv_i\neq0\ {\rm mod\ }\IZ$. If this is not 
the case, some of the sine- or cosine-factors have to be replaced by the 
appropriate volume factors. We find:
\begin{itemize}
\item if $kv_3=0$:
\beq  \label{tadklein3}
\cK_0(k)+\cK_1(k)\ =\ \frac{V_3}{8\tilde s_1\tilde s_2} 
                       [32\,(c_1c_2 + \epsilon\, s_1 s_2)]^2 .
\eeq

\item if $kv_{i\neq3}=0$:
\beq  \label{tadklein12}
\cK_0(k)+\cK_1(k)\ =\ -\frac{V_i}{8\tilde s_3^2} [32\, c_3^2]^2
                - \frac{2}{V_i} [8\,\epsilon]^2
                + \frac{1}{\tilde s_3} [32\,c_3^2][8\,\epsilon] .
\eeq

\item if $kv_3=\pm\half$:
\beq  \label{tadklein3'}
\cK_0(k)+\cK_1(k)\ =\ \frac{2}{V_3} [8\,(1+\epsilon)]^2 .
\eeq

\item if $kv_{i\neq3}=\pm\half$:
\beq  \label{tadklein12'}
\cK_0(k)+\cK_1(k)\ =\ \frac{V_i}{8\tilde s_3^2} [32\, c_3^2]^2
                + \frac{2}{V_i} 8^2
                + \frac{1}{\tilde s_3} [32\,c_3^2][8\,\epsilon] .
\eeq

\item if $k=0$:
\beq \label{tadkleinuntw}
\cK_0(0)+\cK_1(0)\ =\ \frac{V_1V_2V_3}{8} 32^2 
            + \frac{2V_3}{V_1V_2} [32\,\epsilon]^2 
            - V_3 [32\,\epsilon]\cdot 32 .
\eeq
\end{itemize}

The M\"obius strip contribution can be split into the contributions from the
$D9$-branes and from the $D5_3$-branes:
\bea  \label{mobius_ampl}
\cM_9(k) &=  &- 8 \frac{1}{8s_1s_2s_3}  
        \Tr\left(\gamma^{-1}_{\Omega{k},9}\gamma^\top_{\Omega{k},9}\right),\\
\cM_5(k) &=  &- 8 \frac{2 c_1 c_2}{s_3}\,\alpha\Omega^2|_{95_3} 
        \Tr\left(\gamma^{-1}_{\Omega{k},5}\gamma^\top_{\Omega{k},5}\right),
\nonumber
\eea
where $\Omega^2|_{95_3}$ is defined in (\ref{Om95}). There is a sign
ambiguity in the M\"obius strip amplitude. The factor $\alpha\Omega^2|_{95_3}$
in $\cM_5$ has been introduced to reproduce the known tadpoles of the
supersymmetric models of GP and DPBZ \cite{gp,dp,bz} and the \nonsusy models
of AAADS and AU \cite{aaads,au}.
The above equations can be rewritten as
\bea
\cM_9 &= &- 2 \sum_{k=0}^{N/2-1} \frac{c_9}{8\tilde s_1\tilde s_2\tilde s_3}  
              [32\,(c_1c_2c_3 - \mu_9 s_1 s_2 c_3)]  [\Tr{\gamma_{2k,9}}],\\ 
\cM_5 &= &- 2 \sum_{k=0}^{N/2-1} \frac{-\tilde c_5}
              {8\tilde s_1\tilde s_2\tilde s_3}
              [32\,(c_1c_2c_3 - \mu_5 s_1 s_2 c_3)]  
          [4\,\tilde{s}_1 \tilde{s}_2 \Tr{\gamma_{2k,5}}], 
\nonumber
\eea
where we defined $\tilde c_5=-c_5\,\alpha\Omega^2|_{95_3}$. 
The sign $\mu_p$ depends on whether the model has vector structure or not:
$(\gamma_{1,p})^N=\mu_p\,\one$. 
The sign $c_p$ tells us if $\gamma_{\Omega,p}$ is symmetric or antisymmetric:
$\gamma_{\Omega,p}^\top=c_p\,\gamma_{\Omega,p}$.
Again, these formulae are valid only if $2kv_i\neq0\ {\rm mod\ }\IZ$. Else
one has to include the appropriate volume factors. We find:
\begin{itemize}
\item if $kv_3=0$:
\beqa  \label{tadmoeb3}
\cM_9(k) &= &\frac{V_3}{8\tilde s_1\tilde s_2} 
                    2 [\Tr\gamma_{2k,9}] 
                      [-32\,c_9\,(c_1c_2 - \mu_9\, s_1 s_2)],\\
\cM_5(k) &= &\frac{V_3}{8\tilde s_1\tilde s_2} 
                    2 [4\,\tilde s_1\tilde s_2\Tr\gamma_{2k,5}] 
                      [32\,\tilde c_5\,(c_1c_2 - \mu_5\, s_1 s_2)].
\nonumber
\eeqa

\item if $kv_{i\neq3}=0$:
\beqa  \label{tadmoeb12}
\cM_9(k) &= &-\frac{V_i}{8\tilde s_3^2} 
                2 [\Tr\gamma_{2k,9}] [-32\,c_9\,c_3^2]^2
                + \frac{1}{\tilde s_3} [\Tr\gamma_{2k,9}] [8\,c_9\,\mu_9],\\
\cM_5(k) &= & - \frac{2}{V_i} 2 [\Tr\gamma_{2k,5}] [-8\,\tilde c_5\,\mu_5]
            + \frac{1}{\tilde s_3} [\Tr\gamma_{2k,5}] [32\,\tilde c_5\,c_3^2].
\nonumber
\eeqa

\item if $kv_3=\pm\half$:
\beqa  \label{tadmoeb3'}
\cM_9(k) &= &\frac{1}{\tilde s_1} [\Tr\gamma_{2k,9}] [\mp 8\,c_9\,(1-\mu_9)],\\
\cM_5(k) &= &\frac{1}{\tilde s_1} [4\,\tilde s_1\tilde s_2\Tr\gamma_{2k,5}] 
                                  [\pm 8\,\tilde c_5\,(1-\mu_5)].
\nonumber
\eeqa

\item if $kv_{i\neq3}=\pm\half$:
\beqa  \label{tadmoeb12'}
\cM_9(k) &= &\frac{V_i}{8\tilde s_3^2} 
                  2 [\Tr\gamma_{2k,9}] [\pm 32\,c_9\,\mu_9\,c_3^2]
                + \frac{1}{\tilde s_3} [\Tr\gamma_{2k,9}] [\mp 8\,c_9],\\
\cM_5(k) &= &\frac{2}{V_i} 2 [\Tr\gamma_{2k,5}] [\pm 8\,\tilde c_5]
         + \frac{1}{\tilde s_3} [\Tr\gamma_{2k,5}] 
                                [\mp 32\,\tilde c_5\,\mu_5\,c_3^2].
\nonumber
\eeqa

\item if $k=0$:
\beqa \label{tadmoebuntw}
\cM_9(0) &= &\frac{V_1V_2V_3}{8} 2 [\Tr\gamma_{0,9}] [-32\,c_9],\\
\cM_5(0) &= &\frac{2V_3}{V_1V_2} 2 [\Tr\gamma_{0,5}] [-32\,\tilde c_5\,\mu_5].
\nonumber
\eeqa
\end{itemize}

Finally, we want to write the sum of all contributions in a factorised form:
\be  \label{fac_form}
\cC+\cM+\cK\ =\ \sum_{{\rm odd\ }k}[\ldots]^2 + \sum_{k=0}^{N/2-1}[\ldots]^2.
\ee

From the above equations, we see that this factorisation is only possible if
\beq \label{vs}
\epsilon = -\mu_9 = -\mu_5.
\eeq
This means that models with vector structure are only possible if one
uses the alternative $\Omega$-projection of DPBZ (corresponding to 
$\epsilon=-1$). Assuming these conditions, the tadpoles take the following
form:

a) untwisted sector
\be  \label{t_untw}
   \frac{V_1V_2V_3}{8} [\Tr\gamma_{0,9}-32\,c_9]^2 
   + \frac{2V_3}{V_1V_2} [\Tr\gamma_{0,5}-32\,\tilde c_5\,\mu_5]^2 
   + V_3 [\alpha\Tr\gamma_{0,9}\Tr\gamma_{0,5}-32^2\,\epsilon].
\ee

b) twisted sectors without fixed tori, \ie $kv_i\ne0\ {\rm mod\ }\IZ$:
\begin{itemize}
\item odd $k$:
\be  \label{t_oddk}  
   \frac{1}{8s_1s_2s_3} 
        \left[\Tr\gamma_{k,9} + 4\,\alpha\,s_1 s_2\Tr\gamma_{k,5}\right]^2.
\ee

\item even $k=2k'$:
\be  \label{t_evenk}  
   \frac{1}{8\tilde s_1\tilde s_2\tilde s_3} 
        \left[\Tr\gamma_{2k',9} 
              + 4\,\alpha\,\tilde s_1 \tilde s_2\Tr\gamma_{2k',5}
             -32\,c_9\,(c_1c_2c_3 + \epsilon\, s_1 s_2 c_3)\right]^2,
\ee
where $s_i$, $c_i$, $\tilde s_i$ are evaluated with the argument $k'$.
\end{itemize}

c) twisted sectors with fixed tori, \ie $kv_i=0\ {\rm mod\ }\IZ$:
\begin{itemize}
\item odd $k$:
  \begin{itemize}
  \item[-] $i=3$:
  \be  \label{t_fixodd}
      \frac{V_3}{8s_1s_2} 
         \left[\Tr\gamma_{k,9} + 4\,\alpha\,s_1 s_2\Tr\gamma_{k,5}\right]^2.
  \ee

  \item[-] $i\ne3$: \qquad never happens
  \end{itemize}

\item even $k=2k'$, with $k'v_i=0$:
  \begin{itemize}
  \item[-] $i=3$:
  \be  \label{t_fixeven1}  
      \frac{V_3}{8\tilde s_1\tilde s_2} 
         \left[\Tr\gamma_{2k',9} 
               + 4\,\alpha\,\tilde s_1\tilde s_2\Tr\gamma_{2k',5}
         -32\,c_9\,(c_1c_2 + \epsilon\, s_1 s_2)\right]^2.
  \ee

  \item[-] $i\ne3$:
  \bea  \label{t_fixeven2}  
      && \hspace*{-2cm}
         -\frac{V_i}{8\tilde s_3^2} [\Tr\gamma_{2k',9}+32\,c_9\,c_3^2]^2
               - \frac{2}{V_i} [\Tr\gamma_{2k',5}+8\,\tilde c_5\,\,\mu_5]^2
          \nonumber\\
      && + \frac{\alpha}{\tilde s_3} [\Tr\gamma_{2k',9}+32\,c_9\,c_3^2]
                                    [\Tr\gamma_{2k',5}+8\,\tilde c_5\,\,\mu_5].
  \eea
  \end{itemize}

\item even $k=2k'$, with $k'v_i=\pm\half$:
  \begin{itemize}
  \item[-] $i=3$:
  \bea  \label{t_fixeven3}  
     && \hspace*{-2cm}
        \frac{V_3}{8\tilde s_1\tilde s_2} 
         \left[\Tr\gamma_{2k',9} 
               + 4\,\alpha\,\tilde s_1\tilde s_2\Tr\gamma_{2k',5}\right]^2
         +\frac{2}{V_3} [8\,(1+\epsilon)]^2 \nonumber\\
     && \mp\frac{c_9}{\tilde s_1}[\Tr\gamma_{2k',9}
        + 4\,\alpha\,\tilde s_1\tilde s_2\Tr\gamma_{2k',5}] [8\,(1+\epsilon)]. 
  \eea

  \item[-] $i\ne3$:
  \bea  \label{t_fixeven4}  
    && \hspace*{-2cm}
       \frac{V_i}{8\tilde s_3^2} [\Tr\gamma_{2k',9}\pm32\,c_9\,\mu_9\,c_3^2]^2
          + \frac{2}{V_i} [\alpha\,\Tr\gamma_{2k',5}\pm8\,\tilde c_5]^2 
    \nonumber\\
    && + \frac{\alpha}{\tilde s_3} 
             [\Tr\gamma_{2k',9}\pm32\,c_9\,\mu_9\,c_3^2]
             [\Tr\gamma_{2k',5}\pm8\,\tilde c_5].
  \eea
  \end{itemize}
\end{itemize}

From the form of the untwisted tadpoles (\ref{t_untw}), we find that a 
cancellation is only possible if
\be  \label{untw_constr}
c_9=1,\qquad \epsilon=\alpha.
\ee
The first condition could be evaded by introducing anti-$D9$-branes.
Indeed, when fixing the sign of the M\"obius strip amplitude, eq.\
(\ref{mobius_ampl}), we implicitly assumed that anti-branes may only
appear in the 5-brane sector. However, the second condition in 
(\ref{untw_constr}) is always valid. It means that $D$-branes and
$O$-planes must have opposite charges.

\end{appendix}


\newpage

\begin{thebibliography}{99}
\bibitem{bl}
M.~Berkooz, R.~G.~Leigh,
\emph{A $D=4$ $N=1$ Orbifold of Type I Strings},
\NPB{483}{97}{187},
hep-th/9605049.  
%
\bibitem{abpss} 
C.~Angelantonj, M.~Bianchi, G.~Pradisi, A.~Sagnotti, Y.~S.~Stanev, 
\emph{Chiral asymmetry in four-dimensional open-string vacua},
Phys.\ Lett.\ B385 (1996) 96,
hep-th/9606169.
%
\bibitem{ks1}
Z.~Kakushadze, G.~Shiu, 
\emph{A chiral $N=1$ Type I vacuum in four dimensions and its 
      Heterotic dual}, 
Phys.\ Rev.\ D56 (1997) 3686, 
hep-th/9705163.
%
\bibitem{ks2} 
Z.~Kakushadze, G.~Shiu, 
\emph{4-D chiral $N=1$ Type I vacua with and without D5-branes}, 
Nucl.\ Phys.\ B520 (1998) 75, 
hep-th/9706051.
%
\bibitem{z}
G.~Zwart,
\emph{Four-dimensional $N=1$ $Z_N \times Z_M$ Orientifolds},
\NPB{526}{98}{378},
hep-th/9708040.  
%
\bibitem{afiv}
G.~Aldazabal, A.~Font, L.~E.~Ib\'a\~nez, G.~Violero,
\emph{$D=4$, $N=1$ Type IIB Orientifolds},
\NPB{536}{98}{29},
hep-th/9804026.  
%
\bibitem{kst} 
Z.~Kakushadze, G.~Shiu and S.-H.~Tye, 
\emph{Type IIB orientifolds, F-theory, Type I strings on orbifolds and 
      Type I - heterotic duality}, 
Nucl.\ Phys.\ B533 (1998) 25, 
hep-th/9804092.
%
\bibitem{kcpw}
Z.~Kakushadze,
\emph{A Three-Family SU(6) Type I Compactification},
Phys.\ Lett.\ B434 (1998) 269,
hep-th/9804110;
M.~Cvetic, M.~Plumacher, J.~Wang,
\emph{Three Family Type IIB Orientifold String Vacua with Non-Abelian 
      Wilson Lines},
JHEP 0004 (2000) 004,
hep-th/9911021.
%
\bibitem{bgk}
R.~Blumenhagen, L.~G\"orlich, B.~K\"ors,
\emph{Supersymmetric 4D Orientifolds of Type IIA with D6-branes at Angles},
JHEP 0001 (2000) 040,
hep-th/9912204.
%
\bibitem{bi}
J.~D.~Blum, K.~Intriligator,
\emph{New Phases of String Theory and 6d Fixed Points via Branes at Orbifold 
      Singularities},
Nucl.\ Phys.\ B506 (1997) 199,
hep-th/9705044.
%
\bibitem{kr1}
M.~Klein, R.~Rabad\'an,
\emph{Orientifolds with discrete torsion},
JHEP 0007 (2000) 040, 
hep-th/0002103.
%
\bibitem{blpssw}
M.~Berkooz, R.~G.~Leigh, J.~Polchinski, J.~H.~Schwarz, N.~Seiberg, E.~Witten,
\emph{Anomalies, Dualities, and Topology of $D=6$ $N=1$ Superstring Vacua},
\NPB{475}{96}{115},
hep-th/9605184.      
%
\bibitem{bs}
M.~Bianchi, A.~Sagnotti,
\emph{Twist symmetry and open-string Wilson lines},
Nucl.\ Phys.\ B361 (1991) 519.
%
\bibitem{ps}
G.~Pradisi, A.~Sagnotti,
\emph{Open string orbifolds},
Phys.\ Lett.\ B216 (1989) 59;
M.~Bianchi, A.~Sagnotti,
\emph{On the systematics of open-string theories},
Phys.\ Lett.\ B247 (1990) 517.
%
\bibitem{gp}
E.~G.~Gimon, J.~Polchinski,
\emph{Consistency Conditions for Orientifolds and D-Manifolds},
\PRD{54}{96}{1667},
hep-th/9601038.
%
\bibitem{dp}
A.~Dabholkar, J.~Park,
\emph{A Note on Orientifolds and F-theory},
\PLB{394}{97}{302},
hep-th/9607041.  
%
\bibitem{bz}
J.~D.~Blum, A.~Zaffaroni,
\emph{An Orientifold from F Theory},
Phys.\ Lett.\ B387 (1996) 71,
hep-th/9607019.
%
\bibitem{p} 
J.~Polchinski,
\emph{Tensors from K3 Orientifolds}, 
\PRD{55}{97}{6423}, 
hep-th/9606165.
%
\bibitem{kr2}
M.~Klein, R.~Rabad\'an, 
\emph{$Z_N\times Z_M$ orientifolds with and without discrete torsion},
hep-th/0008173.
%
\bibitem{ads}
I.~Antoniadis, E.~Dudas, A.~Sagnotti,
\emph{Brane Supersymmetry Breaking},
Phys.\ Lett.\ B464 (1999) 38,
hep-th/9908023.    
%
\bibitem{au}
G.~Aldazabal, A.~M.~Uranga, 
\emph{Tachyon-free Non-supersymmetric Type IIB Orientifolds via 
      Brane-Antibrane Systems},
JHEP 9910 (1999) 024,
hep-th/9908072. 
%
\bibitem{aiq}
G.~Aldazabal, L.~E.~Ibanez, F.~Quevedo  
\emph{Standard-like Models with Broken Supersymmetry from Type I String Vacua},
JHEP 0001 (2000) 031,
hep-th/9909172. 
%
\bibitem{aaads}
C.~Angelantonj, I.~Antoniadis, G.~D'Apollonio, E.~Dudas, A.~Sagnotti,
\emph{Type I vacua with brane supersymmetry breaking},
Nucl.\ Phys.\ B572 (2000) 36,
hep-th/9911081.     
%
\bibitem{df}
M.~R.~Douglas, 
\emph{D-branes and Discrete Torsion},
hep-th/9807235;
M.~R.~Douglas, B.~Fiol,
\emph{D-branes and Discrete Torsion II},
hep-th/9903031.
%
\bibitem{abiu}
G.~Aldazabal, D.~Badagnani, L.~E~Ib\'a\~nez, A.~M.~Uranga,
\emph{Tadpole versus anomaly cancellation in D=4,6 compact IIB orientifolds},
JHEP 9906 (1999) 031,
hep-th/9904071.
%
\bibitem{lr}
R.~G.~Leigh, M.~Rozali,
\emph{Brane Boxes, Anomalies, Bending and Tadpoles},
Phys.\ Rev.\ D59 (1999) 026004,
hep-th/9807082.
%
\bibitem{ek}
J.~Erler, A.~Klemm,
\emph{Comment on the Generation Number in Orbifold Compactifications},
Commun.\ Math.\ Phys.\ 153 (1993) 579,
hep-th/9207111.
%
\bibitem{iru1}
L.~E.~Ib\'a\~nez, R.~Rabad\'an, A.~M.~Uranga,
\emph{Anomalous U(1)'s in Type I and Type IIB D=4, N=1 string vacua},
Nucl.\ Phys.\ B542 (1999) 112,
hep-th/9808139.
%
\bibitem{klein}
M.~Klein,
\emph{Anomaly cancellation in D=4, N=1 orientifolds and linear/chiral 
      multiplet duality},
Nucl.\ Phys.\ B569 (2000) 362, 
hep-th/9910143.
%
\bibitem{dm}
M.~Douglas, G.~Moore,
\emph{D-branes, Quivers and ALE Instantons},
hep-th/9603167.
%
\bibitem{abk}
I.~Antoniadis, C.~Kounas, C.~Bachas,
Nucl.\ Phys.\ B288 (1987) 87;
I.~Antoniadis, C.~Bachas,
Nucl.\ Phys.\ B298 (1988) 586.
%
\bibitem{bs2}
M.~Bianchi, A.~Sagnotti,
\emph{Open strings and the relative modular group},
Phys.\ Lett.\ B231 (1989) 389.
%
\bibitem{hh}
A.~Hanany, Y.-H.~He,
\emph{Non-Abelian Finite Gauge Theories},
JHEP 9902 (1999) 013,
hep-th/9811183.
%
\bibitem{i}
A.~V.~Sardo-Infirri,
\emph{Resolutions of Orbifold Singularities and Flows on the McKay Quiver},
alg-geom/9610005.
%
\bibitem{pru}
J.~Park, R.~Rabad\'an, A.~Uranga,
\emph{Orientifolding the conifold},
Nucl.\ Phys.\ B570 (2000) 38,
hep-th/9907086.


\end{thebibliography}
\end{document}